\documentclass[journal]{IEEEtran}
\usepackage{amsmath}
\usepackage{url}
\usepackage{amsmath,amssymb,amsfonts}
\usepackage{algorithmic}
\usepackage{graphicx}
\graphicspath{ {./images/} }
\usepackage{textcomp}
\usepackage{color}
\usepackage{array}
\usepackage{float}

\usepackage{booktabs}
\usepackage{amsbsy}
\usepackage{mathrsfs}
\usepackage{multirow}
\usepackage{subfigure}

\usepackage{lipsum}
\usepackage{threeparttable}

\usepackage{pdfpages}
\usepackage{docmute}
\usepackage{subfiles}

\usepackage{cite}

\ifCLASSINFOpdf
\else
\fi
\begin{document}
%
\title{Multi-View Spatial-Temporal Graph Convolutional Networks with Domain Generalization for Sleep Stage Classification}
%
%
%

\author{Ziyu Jia,
        Youfang Lin,
        Jing Wang,
        Xiaojun Ning,
        Yuanlai He,
        Ronghao Zhou,
        Yuhan Zhou,
        Li-wei H. Lehman
        
\thanks{This work was supported by the National Natural Science Foundation of China under Grant 61603029. Li-wei H. Lehman was supported by the NIH grant R01EB030362. Ziyu Jia was supported by the Swarma-Kaifeng Workshop which is sponsored by Swarma Club and Kaifeng Foundation. \textit{(Corresponding author: Jing Wang)}}
\thanks{Ziyu Jia, Youfang Lin, Jing Wang, Xiaojun Ning, Yuanlai He, Ronghao Zhou, and Yuhan Zhou are with the School of Computer and Information Technology, Beijing Jiaotong University, Beijing 100044, China (e-mail: ziyujia@bjtu.edu.cn; yflin@bjtu.edu.cn; wj@bjtu.edu.cn; ningxj@bjtu.edu.cn; yuanlaihe@bjtu.edu.cn; rhzhou@bjtu.edu.cn; yhzhou@bjtu.edu.cn).}
\thanks{Li-wei H. Lehman is with the Institute for Medical Engineering and Science, Massachusetts Institute of Technology, Cambridge, MA, USA (e-mail: lilehman@mit.edu).}}

%
%

\markboth{IEEE Transactions on Neural Systems and Rehabilitation Engineering,  VOL. XX, NO. XX, XX XX}%
{Shell \MakeLowercase{\textit{et al.}}: Multi-View Spatial-Temporal Graph Convolutional Networks with Domain Generalization for Sleep Stage Classification}
%



\maketitle

\begin{abstract}
Sleep stage classification is essential for sleep assessment and disease diagnosis. Although previous attempts to classify sleep stages have achieved high classification performance, several challenges remain open: 1) How to effectively utilize time-varying spatial and temporal features from multi-channel brain signals remains challenging. Prior works have not been able to fully utilize the spatial topological information among brain regions. 2) Due to the many differences found in individual biological signals, how to overcome the differences of subjects and improve the generalization of deep neural networks is important. 3) Most deep learning methods ignore the interpretability of the model to the brain. To address the above challenges, we propose a multi-view spatial-temporal graph convolutional networks (MSTGCN) with domain generalization for sleep stage classification. Specifically, we construct two brain view graphs for MSTGCN based on the functional connectivity and physical distance proximity of the brain regions. The MSTGCN consists of graph convolutions for extracting spatial features and temporal convolutions for capturing the transition rules among sleep stages. In addition, attention mechanism is employed for capturing the most relevant spatial-temporal information for sleep stage classification. Finally, domain generalization and MSTGCN are integrated into a unified framework to extract subject-invariant sleep features. Experiments on two public datasets demonstrate that the proposed model outperforms the state-of-the-art baselines. 
\end{abstract}


%
\IEEEpeerreviewmaketitle

\section{Introduction}
%
%
%
%
\IEEEPARstart{S}{leep} stage classification is important for the assessment of sleep quality and the diagnosis of sleep disorders. Sleep experts identify sleep stages based on American Academy of Sleep Medicine (AASM) standard \cite{berry2012rules} and observations recorded in polysomnography (PSG), which includes electroencephalography (EEG) at different positions on the head and electrooculography (EOG). The transition rules among different sleep stages recorded in the AASM standard, which can assist sleep experts in identifying the sleep stages. Although these rules provide valuable information, classifying the sleep stages by human sleep experts is still a tedious and time-consuming task. Moreover, the classification results are affected by the variability and subjectivity of sleep experts.

Automatic sleep stage classification can greatly improve the efficiency of traditional sleep stage classification and has important clinical value. Many researchers have made great contributions to automate this classification task. At first, traditional machine learning methods based on time domain, frequency domain, and time-frequency domain features are adopted \cite{alickovic2018ensemble, memar2017novel}. However, the classification accuracy of these methods depends heavily on feature engineering and feature selection, which require substantial expert knowledge. Recently, deep learning methods have been widely applied to automatically classify sleep stage thanks to its powerful ability of representation learning. For example, Convolutional Neural Network (CNN) \cite{lecun1998gradient} and Recurrent Neural Network (RNN) \cite{elman1990finding} are often utilized to learn appropriate feature representations from transformed data or directly from raw data.

Although the existing methods \cite{jia2021salientsleepnet,chambon2018deep,cai2020brainsleepnet,jia2020sleepprintnet,jia2020sleep} achieve high accuracy for sleep stage classification, these methods have not sufficiently solved the following challenges: 1)  The spatial-temporal features of sleep stages have not been fully considered. In particular, the topology among brain regions has not been effectively employed to capture richer spatial features. 2) Physiological signals vary significantly across different subjects, which hinders the generalizability of the trained classifiers. 3) Most deep learning methods, especially related graph neural network models, ignore the importance of model interpretability to the brain.

There have been several attempts to address \textit{the first challenge} \cite{dong2017mixed, supratak2017deepsleepnet, chambon2018deep, phan2019seqsleepnet}. For example, CNN is usually applied to extract the spatial features of the brain, and RNN is applied to capture temporal features during sleep transition. However, the limitation of these networks is that their input must be grid data (image-like representations) without utilizing the connections among brain regions \cite{gopinath2019adaptive}. Due to the fact that brain regions are in non-Euclidean space, graph is the most appropriate data structure to indicate brain connection. Therefore, GraphSleepNet \cite{jia2020graphsleepnet} is proposed to classify sleep stages based on the functional connectivity of the brain network and using spatial-temporal graph convolution to achieve the state-of-the-art performance. However, in the brain network based on functional connectivity, there may not necessarily be connections among physically adjacent brain regions. In fact, existing neuroscience research shows that brain regions that are adjacent to each other at physical distances can influence each other \cite{salvador2005neurophysiological}. However, GraphSleepNet only utilizes the functional connectivity of the brain to construct the sleep stage networks, which ignores the importance of the physical proximity of the brain in space. For \textit{the second challenge}, some researchers try to apply transfer learning methods to improve the generalization of the models \cite{phan2020personalized, banluesombatkul2020metasleeplearner}. The existing sleep stage classification models based on transfer learning are all two-step training paradigms. That is, these models need to be pre-trained and then fine-tuned to new subject data. The fine-tuning operation needs to collect sleep data from specific new subjects or datasets, which is quite expensive and inconvenient. In addition, the generalization of transfer learning models that need to be fine-tuned is limited. These models are designed for specific subjects and may not show excellent performance on other new subjects. Therefore, fine-tuning is only applicable to the personalized (subject-variant) model of the specific subject. And whenever a new subject needs to be evaluated, the existing model must be re-collected and re-trained. Therefore, for clinical systems suitable for unknown users, fine-tuning may become inefficient. For \textit{the third challenge}, previous attempts to develop interpretable CNN or RNN classification models have been sparse \cite{supratak2017deepsleepnet, chambon2018deep, sokolovsky2019deep}. Specifically, no attempt has been made to interpret the key modules of graph neural network for sleep stage classification from the perspective of the brain network.

\begin{figure*}[ht]
\centering
\includegraphics[width=0.84\textwidth]{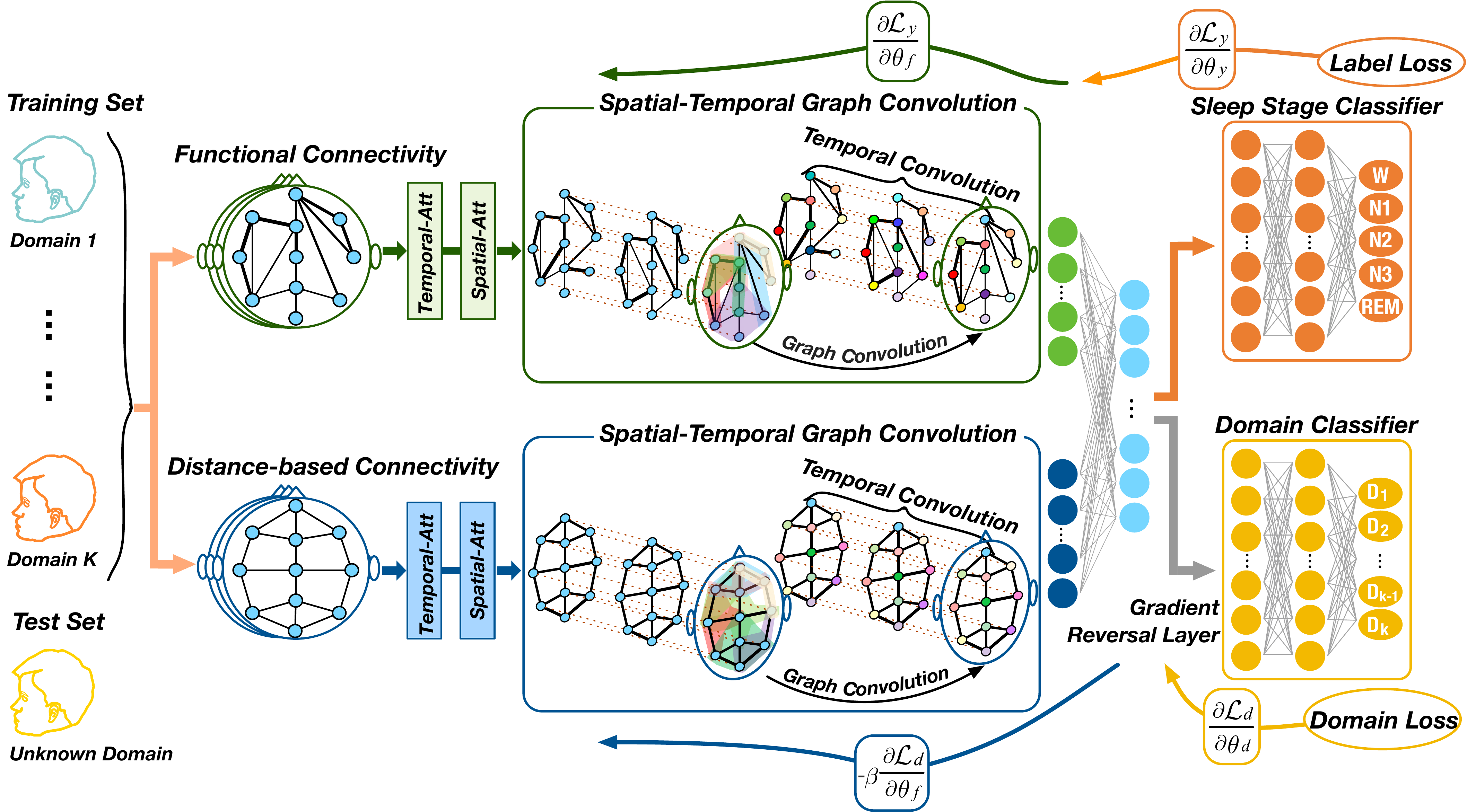} 
\caption{The overall architecture of the MSTGCN for sleep stage classification. First of all, two different views on the brain are constructed: the functional connectivity-based brain graph and spatial distance-based brain graph. Different views reflect different spatial relationships of the brain. Then, an attention based spatial-temporal graph convolution is designed for the most relevant spatial-temporal features for sleep stage classification. Finally,  a domain generalization with the gradient reversal layer is implemented to improve the generalization of the model. In domain generalization, each subject in training set is treated as a specific source  domain. The advantage of domain generalization over other transfer learning methods is that this method does not require any information (a small number of labeled samples or unlabeled sample data distribution) from the test set (called unknown domain or target domain). Therefore, domain generalization improves the generalization of the model and is more suitable for clinical systems applied by unknown users.} 
\label{OA}
\end{figure*}

In order to address the above challenges, we propose the multi-view spatial-temporal graph convolutional networks (MSTGCN) with domain generalization for sleep stage classification. Figure \ref{OA} illustrates the overall architecture of our model. Specifically,  1) we construct two brain view graphs based on the spatial proximity and functional connectivity of the brain, where each EEG channel corresponds to a node of the graph, and the specific connections among the channels correspond to the edge of the graph. 2) Then, we utilize spatial graph convolution to capture rich spatial features. Temporal convolution is applied for capturing the transition among different sleep stages. Actually, sleep experts usually identify the class label of one sleep state according to both the characteristic EEG waves of the current state and the class labels of its neighbors. 3) We design a spatial-temporal attention mechanism to capture the most relevant spatial-temporal information on the sleep stages. 4) Finally, we apply the adversarial domain generalization, which is a typical method of transfer learning without fine-tuning. In the process of model training, each subject is employed as a specific source domain for subject-invariant sleep feature extraction. The subject-invariant sleep feature does not vary with different subjects and is related to sleep stage classification. The advantage of the domain generalization is that it does not require any information in the new subjects (target domain).

To the best of our knowledge, it is the first attempt to apply spatial-temporal graph neural networks with domain generalization for sleep stage classification. Overall, the main contributions of the proposed model for sleep stage classification are summarized as follows:

\begin{itemize}
\item We construct different brain views based on the functional connectivity and physical distance proximity of the brain. The complementarity of different views provides rich spatial topology information for classification tasks.
\item We design a spatial-temporal graph convolution with attention mechanism, which consists of spatial-temporal graph convolution for spatial-temporal features and attention mechanism for capturing the most relevant spatial-temporal information for sleep stage classification. 
\item We integrate domain generalization and spatial-temporal graph convolutional networks into a unified framework to extract subject-invariant sleep features.
\item We conduct experiments on two public sleep datasets, namely ISRUC-S3 and MASS-SS3. Experimental results demonstrate that the proposed model achieves the state-of-the-art performance.
\item We explore the interpretability of the key modules of the model. In particular, we present the functional connectivity obtained through adaptive graph learning. The results indicate that functional connectivity during light sleep is more complex than that during deep sleep. 
\end{itemize}

Compared to the Adaptive Spatial-Temporal Graph Convolutional Networks (called GraphSleepNet) published in our preliminary work \cite{jia2020graphsleepnet}, MSTGCN has the following important improvements:
1) The brain network based on physical distance proximity is constructed. It and the preliminary adaptive functional connectivity brain network form a multi-view brain network, which can provide rich brain spatial topology information for sleep stage classification.
2) Domain generalization is integrated with spatial-temporal graph convolutional networks into a unified framework to improve the generalization of the proposed model.
3) Experiments are conducted to evaluate the effectiveness of MSTGCN on two sleep datasets, of which ISRUC-S3 is not evaluated in our preliminary work. Moreover, we conduct the ablation experiments to evaluate the impact of each component of MSTGCN on the performance.
4) The interpretability of the key modules in MSTGCN is explored and discussed.

\section{Related Work}


In recent years, time series analysis has attracted the attention of many researchers \cite{jia2020refined,jia2019detecting}. As a typical time series, physiological signals are used in many fields, such as motor imagery \cite{li2020learning,ziyu2020motor,10.1007/978-3-030-67664-3_44}, emotion recognition \cite{jia2020sst,jia2021hetemotionnet}, and sleep stage classification \cite{jia2020graphsleepnet}, etc. With the development of deep learning, two popular deep learning models, CNN and RNN, are widely applied in sleep stage classification. Specifically, a fast discriminative complex-valued CNN (FDCCNN) \cite{zhang2017new} is proposed to capture the sleep information hidden inside EEG signals. A CNN model based on multivariate and multimodal physiological signals \cite{chambon2018deep} takes into account the transitional rules of sleep stages to assist classification. 
A hierarchical RNN named SeqSleepNet \cite{phan2019seqsleepnet} tackles the task as a sequence-to-sequence classification task. At the same time, hybrid models are also employed by some researchers. DeepSleepNet \cite{supratak2017deepsleepnet} utilizes CNN to extract time-invariant features, and Bi-directional Long Short-Term Memory (BiLSTM) to learn the transition rules among sleep stages. A hierarchical neural network \cite{sun2019hierarchical} implements comprehensive feature learning stage and sequence learning stage, respectively. Additionally, with the development of attention mechanisms, a deep Bi-directional RNN with attention mechanism is utilized for single-channel sleep staging \cite{phan2018automatic}. 

Although CNN and RNN models achieve high accuracy, their limitation is that the model’s input must be grid data ignoring the connection among brain regions. As different brain regions are not in the Euclidean space, grid data may not be the optimal data representation. Hence, the graph is the most appropriate data structure. GraphSleepNet \cite{jia2020graphsleepnet} is proposed to utilize graph neural network to model functional connectivity brain network to achieve the SOTA performance. However, it only considers the spatial functional connectivity, and to a certain extent ignores the spatial proximity of brain regions.

Some previous researchers attempt to solve the subject difference problem found in physiological signals. Transfer learning methods are applied to improve the robustness of deep learning models for individual differences \cite{phan2020personalized,banluesombatkul2020metasleeplearner}. For example, MetaSleepLearner \cite{banluesombatkul2020metasleeplearner} based on model-agnostic meta-learning is proposed to overcome the subject difference problem by training in the source domain and fine-tuning in the target domain. Although the existing transfer learning methods for sleep stage classification can achieve improved results, almost all existing work needs to fine-tune the pre-trained model for sleep stage classification. That is, these models require additional fine-tuning operations using part of the labeled data in the target domain. Therefore, these transfer learning methods that need to be fine-tuned are only suitable for the specific subject's personalized model. In this case, whenever a new subject needs to be evaluated, data must be collected again and the existing model must be fine-tuned again. Therefore, for clinical systems that need to be adapted to unknown subjects, fine-tuning operations may become inefficient.

\section{Preliminaries}

\textbf{Sleep Stage.}
Polysomnography (PSG) is usually employed for recording physiological signals during sleep in clinical medicine. The PSG is segmented into 30-second epochs for sleep stage classification. Sleep experts usually classify sleep epochs into different stages based on the sleep staging standard. Specifically, according to the AASM sleep staging standard, the human sleep process can be divided into three main parts: Wakefulness (Wake), rapid eye movement (REM), and non-rapid eye movement (NREM). Furthermore, the NREM can be subdivided into three parts: N1 stage, N2 stage, and N3 stage. In general, sleep experts directly divide the sleep state into 5  different classes (Wake, N1, N2, N3, and REM).

\begin{figure}[ht]
\centering
\includegraphics[width=0.95\columnwidth, trim={0 30 0 10},]{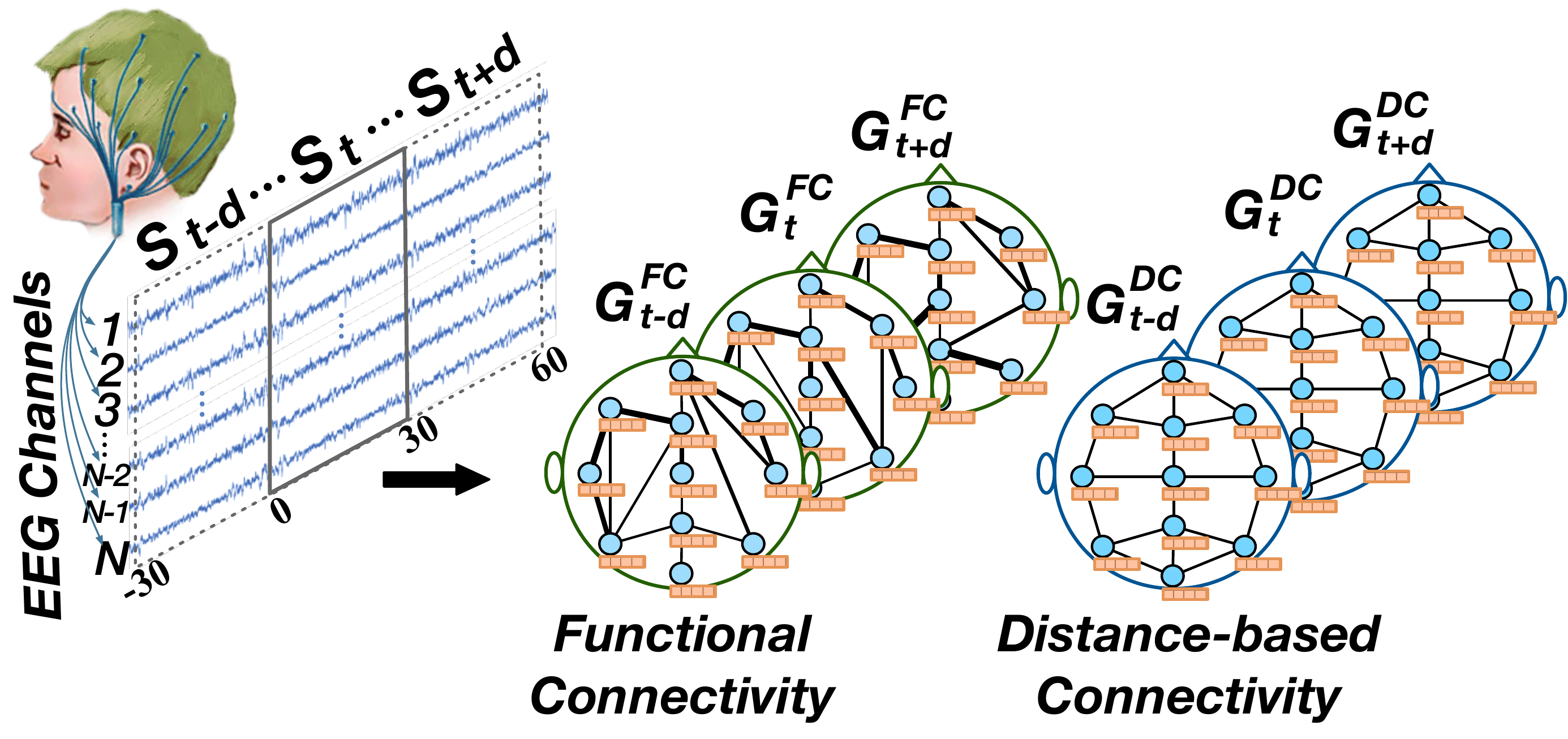}
\caption{Multi-view sleep brain network. Left network is the functional connectivity-based network and right network is the spatial distance-based network.} 
\label{pre}
\end{figure}

\textbf{Sleep Brain Network.}
A sleep brain network is defined as a graph $G=(V,E,\boldsymbol{A})$, where $V$ represents the set of vertices and each vertex in the network represents an electrode on brain; $|V|=N$ is the number of vertices in sleep brain network; $E$ denotes the set of edges and indicates the connection between vertices; $\boldsymbol{A}$ denotes the adjacency matrix of sleep brain network $G$. As presented in Figure \ref{pre}, $G^{FC}_{t}$ represents sleep brain network constructed from the functional connectivity and $G^{DC}_{t}$ represents sleep brain network constructed from spatial distance. And a 30s EEG signal sequence $S_{t}$ (called a sleep epoch) is transformed into $G^{FC}_{t}$  and $G^{DC}_{t}$.

\textbf{Sleep Feature Matrix.}
The sleep feature matrix is the input of the graph neural network. We define the raw signals sequence as $\boldsymbol{S}=\left({S}_{1}, {S}_{2}, \ldots, {S}_{L}\right) \in \mathbb{R}^{N \times T_{s} \times L}$, where $L$ is the number of sleep epochs, $T_s$ represents the time series length of each sleep epoch ${S}_{i} \in \boldsymbol{S} (i \in\{1,2, \cdots, L\})$. For each sleep epoch ${S}_{i}$, we extract the node feature by using a feature extraction network in Supplementary Material S.1 and define each epoch ${S}_{i}$'s feature matrix $\boldsymbol{X}_{i}=\left(\boldsymbol{x}_{1}^{i}, \boldsymbol{x}_{2}^{i}, \ldots, \boldsymbol{x}_{N}^{i}\right)^{\mathrm{T}} \in \mathbb{R}^{N \times F_{d}}$, where $\boldsymbol{x}_{n}^{i} \in \mathbb{R}^{F_{d}} (n \in \{1,2, \cdots, N\})$ represents $F_{d}$ features of node $n$ at epoch $i$. 

\textbf{Sleep Stage Classification Problem.}
The research goal is to learn the mapping relationship between the encoded signals and sleep stage classes. The problem of sleep stage classification is defined as: given $\boldsymbol{\mathcal{S}}=\left({S}_{i-d}, \ldots, {S}_{i}, \ldots, {S}_{i+d}\right) \in \mathbb{R}^{N \times T_{s} \times T_{n}}$ identify the current sleep stage $y$, where $\boldsymbol{\mathcal{S}}$ represents the temporal context of ${S}_i$, $y$ denotes the ${S}_i$'s sleep stage class label, and $T_n=2d+1$ is the number of sleep brain networks, where $d \in \mathbb{N}^+$ is temporal context.
Specifically, in order to identify the sleep stage of the current sleep epoch ${S}_{i}$, we utilize its previous $d$ epochs and following $d$ epochs as the context. For each epoch, we construct $G^{DC}$ and $G^{FC}$ respectively, and they are employed as the input of our model to identify the sleep stage $y$ of the current sleep epoch.

\section{Multi-View Spatial-Temporal GCN}
The overall architecture of the proposed model is exhibited in Figure \ref{OA}. We summarize four key ideas of the proposed MSTGCN model: 1) Construct multiple views of the brain connection to fully indicate the spatial information of the brain. 2) Combine spatial graph convolution and temporal convolution to extract both spatial and temporal features. 3) Employ a spatial-temporal attention mechanism to automatically pay more attention to valuable spatial-temporal information. 4) Integrate domain generalization and spatial-temporal GCN in a unified framework to extract subject-invariant sleep features. The overall architecture is designed to accurately identify sleep stages.

\subsection{Multi-view on Brain Graph}
In this section, we introduce two different views from the brain graph: the functional connectivity-based brain graph and spatial distance-based brain graph. Different views reflect different spatial relationships of the brain. Specifically, the functional connectivity-based brain graph can present the collaboration of different brain regions in space. The actual physical locations of these brain regions may not be adjacent. 
However, existing neuroscience studies have presented physically adjacent brain regions also interact.
Therefore, these two views on brain have a certain degree of complementarity and can fully demonstrate the spatial relationship of the brain.

\subsubsection{Functional Connectivity-based Brain Graph}
\begin{figure}[htb]
    \centering
    \includegraphics[width=0.75\columnwidth, trim={0 30 0 11.5},]{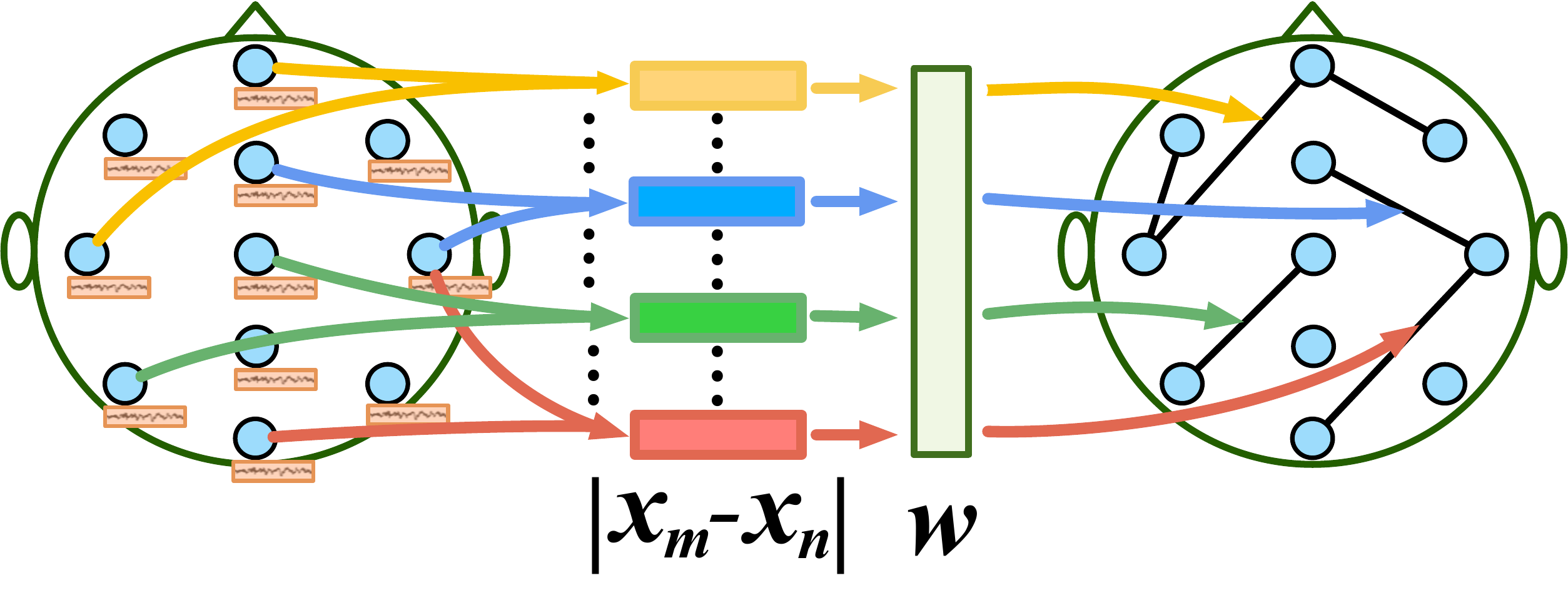}
    \caption{The adaptive sleep graph learning to generate functional connectivity for sleep stage classification. $\boldsymbol{x}_m$ and $\boldsymbol{x}_n$ represent the features of two nodes respectively, $\boldsymbol{w}$ is learnable weight. The more similar the node features, the greater the probability of establishing a connection.}
    \label{GL}
\end{figure}

Functional connectivity is usually constructed based on correlations or dependencies among physiological signals \cite{tagliazucchi2012automatic}. Pearson Correlation Coefficient (PCC) \cite{pearson1903laws} and Mutual Information (MI) \cite{danon2005comparing} are two common methods to determine the functional connectivity of the brain. Due to the limited understanding of the brain, it is still challenging to determine a suitable graph structure in advance for sleep stage classification. Hence, we propose a data-driven graph generation for functional connectivity. This data-driven approach constructs the functional connectivity graphs adaptively for different sleep stages based on the feature correlation between nodes as displayed in Figure \ref{GL}. 
We define a non-negative function $A_{m n}^{FC}=g\left(\boldsymbol{x}_{m}, \boldsymbol{x}_{n}\right) (m,n \in\{1,2, \cdots, N\})$ to represent the functional connectivity between nodes $\boldsymbol{x}_{m}$ and $\boldsymbol{x}_{n}$ based on the input feature matrix $\boldsymbol{X}_{i}=\left(\boldsymbol{x}_{1}^{i}, \boldsymbol{x}_{2}^{i}, \ldots, \boldsymbol{x}_{N}^{i}\right)^{\mathrm{T}} \in \mathbb{R}^{N \times F_{d}}$.
$g\left(\boldsymbol{x}_{m}, \boldsymbol{x}_{n}\right)$ is implemented through a layer neural network, which has the learnable weight vector $\boldsymbol{w}=\left(w_{1}, w_{2}, \ldots, w_{F_{d}}\right)^{\mathrm{T}} \in \mathbb{R}^{F_{d} \times 1}$. The learned graph structure (adjacency matrix) $\boldsymbol{A}^{FC}$ is defined as:
\begin{equation}
A_{m n}^{FC} \! = \! g(\boldsymbol{x}_{m},\! \boldsymbol{x}_{n})\! =\! \frac{\exp (\operatorname{ReLU}(\boldsymbol{w}^{\mathrm{T}}|\boldsymbol{x}_{m}\! -\! \boldsymbol{x}_{n}|))}{\sum_{n=1}^{N} \exp \left(\operatorname{ReLU}\left(\boldsymbol{w}^{\mathrm{T}}\left|\boldsymbol{x}_{m}\! -\! \boldsymbol{x}_{n}\right|\right)\right)}
\end{equation}
where rectified linear unit (ReLU) is an activation function to guarantee that $A_{m n}^{FC}$ is non-negative. The softmax operation normalizes each row of $\boldsymbol{A}^{FC}$.
The weight vector $\boldsymbol{w}$ is updated by minimizing the following loss function,
\begin{equation}
\mathcal{L}_{\text{graph\_learning}}=\sum_{m, n=1}^{N}\left\|\boldsymbol{x}_{m}-\boldsymbol{x}_{n}\right\|_{2}^{2} A_{m n}^{FC}+\lambda\|\boldsymbol{A}^{FC}\|_{F}^{2}
\label{GLL}
\end{equation}
That is, the larger distance $\left\|\boldsymbol{x}_{m}-\boldsymbol{x}_{n}\right\|_{2}$ between $\boldsymbol{x}_m$ and $\boldsymbol{x}_n$, the smaller $A_{m n}^{FC}$ is. Due to the brain connection structure is not a fully connected graph, we utilize the second term in the loss function to control the sparsity of graph $\boldsymbol{A}^{FC}$, where $\lambda =0.001$ is a regularization parameter.

The proposed graph generation mechanism automatically constructs the neighborhood connection of the nodes. To avoid the trivial solution (i.e., $\boldsymbol{w}=(0,0, \cdots, 0)$), which is due to minimizing the above loss function $\mathcal{L}_{\text{graph\_learning}}$ independently, we utilize it as a regularized term to form the loss function.

\subsubsection{Spatial Distance-based Brain Graph}
\begin{figure}[htb]
    \centering
    \includegraphics[height=21mm]{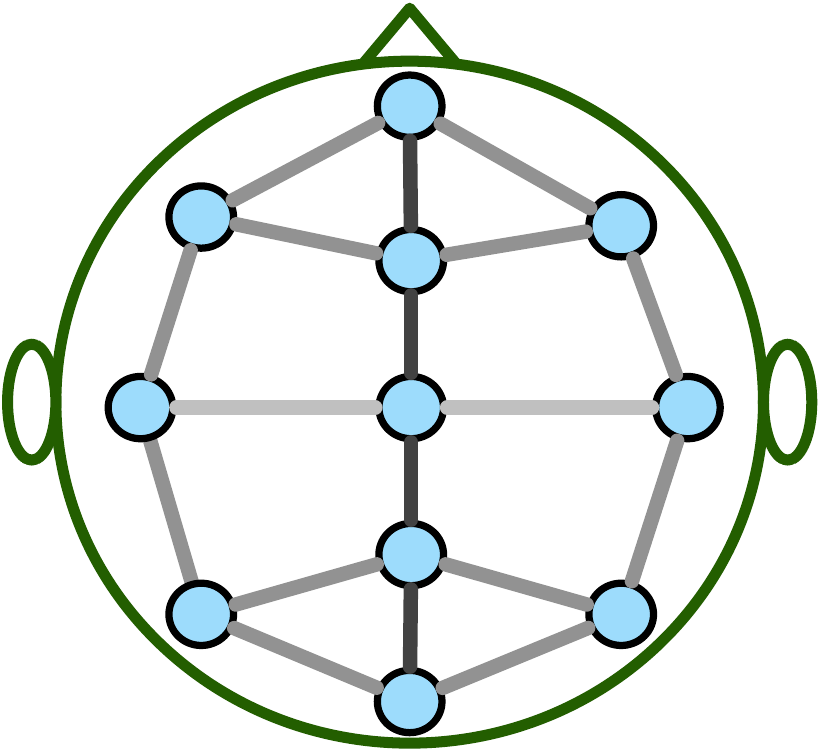}
    \caption{The spatial distance-based brain graph for sleep stage classification.}
    \label{DC1}
\end{figure}

Previous studies have presented that adjacent brain regions affect each other and the strength of the impact is inversely proportional to the actual physical distance \cite{salvador2005neurophysiological}. That is, the closer the distance between brain regions, the greater the impact. Therefore, we construct a spatial distance-based brain graph for sleep stage classification, as illustrated in Figure \ref{DC1}.

\subsection{Spatial-Temporal Attention}
The attention mechanism is often utilized to automatically extract the most relevant information. In this study, we employ a spatial-temporal attention mechanism \cite{jia2020graphsleepnet} to capture valuable spatial-temporal information on the sleep brain network. The spatial-temporal attention mechanism contains spatial attention and temporal attention.

\subsubsection{Spatial Attention}
In the spatial dimension, different regions have different effects on the sleep stage which are dynamically changing during sleep. To automatically extract the attentive spatial dynamics, we utilize a spatial attention mechanism, which is defined as follows (take the spatial attention based on the functional connectivity view as an example):
\begin{equation}
\boldsymbol{P}\! =\! \boldsymbol{V}_{p} \cdot \sigma((\boldsymbol{\mathcal{X}}^{(l-1)} \boldsymbol{Z}_{1}) \boldsymbol{Z}_{2}(\boldsymbol{Z}_{3} \boldsymbol{\mathcal{X}}^{(l-1)})^{\mathrm{T}}\! +\! \boldsymbol{b}_{p})
\end{equation} 
\begin{equation}
  P_{m, n}^{\prime}=\operatorname{softmax}(P_{m, n})
\end{equation}
where $\boldsymbol{\mathcal{X}}^{(l-1)}=\left(\boldsymbol{X}_{1}, \boldsymbol{X}_{2}, \ldots, \boldsymbol{X}_{T_{l-1}}\right) \in \mathbb{R}^{N \times C_{l-1} \times T_{l-1}}$ is the $l$-th layer's input. $C_{l-1}$ represents neural network channel's number of each node, i.e., $l=1, \quad C_{0}=F_{d}$. $T_{l-1}$ denotes the $l$-th layer's temporal dimension.
$\boldsymbol{V}_{p}, \boldsymbol{b}_{p}\in \mathbb{R}^{N \times N}$, $\boldsymbol{Z}_{1} \in \mathbb{R}^{T_{l-1}}$, $\boldsymbol{Z}_{2} \in \mathbb{R}^{C_{l-1} \times T_{l-1}}$, $\boldsymbol{Z}_{3} \in \mathbb{R}^{C_{l-1}}$ are learnable parameters, $\sigma$ denotes the sigmoid activation function.
$\boldsymbol{P}$ represents spatial attention matrix, which is dynamically computed by current layer's input. $P_{m,n}$ represents the correlation between node $m$ and $ n$. The softmax operation is utilized to normalize the attention matrix $\boldsymbol{P}$. In the proposed model, when the graph convolution is performed, the learned adjacency matrix  $\boldsymbol{A}^{FC}$ and spatial attention matrix $\boldsymbol{P}$ can dynamically adjust the update of nodes.

\subsubsection{Temporal Attention}
In the temporal dimension, there are correlations among neighboring sleep stages, and the correlations vary in different situations. Therefore, a temporal attention mechanism is utilized to capture dynamic temporal information among sleep brain networks.

The temporal attention mechanism is defined as follows:
\begin{equation}
\boldsymbol{Q}=\boldsymbol{V}_{q} \cdot \sigma(((\boldsymbol{\mathcal{X}}^{(l-1)})^{\mathrm{T}} \boldsymbol{M}_{1}) \boldsymbol{M}_{2}(\boldsymbol{M}_{3} \boldsymbol{\mathcal{X}}^{(l-1)})+\boldsymbol{b}_{q})
\end{equation}
\begin{equation}
    Q_{u, v}^{\prime}=\operatorname{softmax}(Q_{u, v})
\end{equation}
where $\boldsymbol{V}_{q}, \boldsymbol{b}_{q} \in \mathbb{R}^{T_{l-1} \times T_{l-1}}$, $\boldsymbol{M}_{1} \in \mathbb{R}^{N}$, $\boldsymbol{M}_{2} \in \mathbb{R}^{C_{l-1} \times N}$, $\boldsymbol{M}_{3} \in \mathbb{R}^{C_{l-1}}$ denotes learnable parameters. $Q_{m, n}$ denotes the strength of correlation between sleep brain network $G_u$ and $G_v$. Finally, the softmax operation is utilized to normalize the attention matrix $\boldsymbol{Q}$. 
 The input of the MST-GCN is tuned by the temporal attention:
 $\boldsymbol{\hat{\mathcal{X}}}^{(l-1)}=(\boldsymbol{\hat{X}}_{1}, \boldsymbol{\hat{X}}_{2}, \ldots, \boldsymbol{\hat{X}}_{T_{l-1}})=(\boldsymbol{X}_{1}, \boldsymbol{X}_{2}, \ldots, \boldsymbol{X}_{T_{l-1}})\boldsymbol{Q}^{\prime} \in \mathbb{R}^{N \times C_{l-1} \times T_{l-1}}$
 to pay more attention to informative temporal information.

\subsection{Spatial-Temporal Graph Convolution}
Spatial-temporal graph convolution is a combination of spatial graph convolution and standard temporal convolution, which is utilized to extract both spatial and temporal features. The spatial features are extracted by aggregating information from neighbor nodes for each sleep brain network and the temporal features are captured by exploiting temporal dependencies from neighbor sleep stages.

\subsubsection{Spatial Graph Convolution}
We employ graph convolution based on spectral graph theory to extract spatial features in the spatial dimension. For each sleep stage to be identified, the adjacency matrices $\boldsymbol{A}^{FC}$ and $\boldsymbol{A}^{DC}$ are provided for graph convolution. 
In addition, we employ the Chebyshev expansion of graph Laplacian to reduce computational complexity.
Chebyshev graph convolution \cite{defferrard2016convolutional} using the $K-1$ order polynomials is defined as:
\begin{equation}
g_{\theta} *_{G} x=g_{\theta}(\mathbf{L}) x=\sum_{k=0}^{K-1} \theta_{k} T_{k}(\tilde{\mathbf{L}}) x
\end{equation}
where $g_{\theta}$ denotes the convolution kernel, $*_G$ denotes the graph convolution operation, $\theta \in \mathbb{R}^{K}$ is a vector of Chebyshev coefficients and $x$ is the input data. $\boldsymbol{L}=\boldsymbol{D}-\boldsymbol{A}$ is Laplacian matrix, where $\boldsymbol{D} \in \mathbb{R}^{N \times N}$ is degree matrix. $\tilde{\boldsymbol{L}}=\frac{2}{\lambda_{\max }} \boldsymbol{L}-\boldsymbol{I}_{N}$, where $\lambda_{\max }$ is Laplacian matrix's maximum eigenvalue and $\boldsymbol{I}_{N}$ is an identity matrix. 
$T_{k}(x)$ is the Chebyshev polynomials recursively.

The information of neighboring 0 to $K - 1$ order neighbors centered at each node is extracted via the approximate expansion of Chebyshev polynomial.

We generalize the above definition to the nodes with multiple neural network channels. The $l$-th layer's input is $\boldsymbol{\hat{\mathcal{X}}}^{(l-1)}=(\boldsymbol{\hat{X}}_{1}, \boldsymbol{\hat{X}}_{2}, \ldots, \boldsymbol{\hat{X}}_{T_{l-1}}) \in \mathbb{R}^{N \times C_{l-1} \times T_{l-1}}$, where $C_{l-1}$ represents neural network channel's number of each node, $T_{l-1}$ denotes the $l$-th layer's temporal dimension.
For each $\boldsymbol{\hat{X}}_{i}$, we obtain $g_{\theta} *_{G} \boldsymbol{\hat{X}}_{i}$ by using $C_l$ filters on $\boldsymbol{\hat{X}}_i$, where $\Theta=\left(\Theta_{1}, \Theta_{2}, \ldots, \Theta_{C_{l}}\right) \in \mathbb{R}^{K \times C_{l-1} \times C_{l}}$ is the convolution kernel parameter \cite{defferrard2016convolutional}.
Hence, the information of the $0 \sim K-1$ order neighbors is aggregated to each node.
\subsubsection{Temporal Convolution}
To capture the sleep transition rules, which are utilized by sleep experts to classify the current sleep stage in combination with neighboring sleep stages, we employ CNN to perform convolution operation in the temporal dimension. Specifically, after graph convolution operation has sufficiently extracted the spatial features from each sleep brain network, we implement a standard 2D convolution layer to extract the temporal context information of the current sleep stage. The temporal convolution operation on the $l$-th layer is defined as:
\begin{equation}
\boldsymbol{\mathcal{X}}^{(l)}=\operatorname{ReLU}(\Phi *(\operatorname{ReLU}(g_{\theta} *_{G} \boldsymbol{\hat{\mathcal{X}}}^{(l-1)}))) \in \mathbb{R}^{N \times C_{l} \times T_{l}}
\end{equation}
where ReLU is the activation function, $\Phi$ denotes the convolution kernel's parameters, * denotes the standard convolution operation.

After the multi-view ST-GCN extracts a large number of features, we employ the concatenate operation to perform feature fusion on $\boldsymbol{\mathcal{X}}^{FC}$ and $\boldsymbol{\mathcal{X}}^{DC}$:
\begin{equation}
    \boldsymbol{\mathcal{X}} = \boldsymbol{\mathcal{X}}^{FC} \parallel \boldsymbol{\mathcal{X}}^{DC} 
\end{equation}
where $\boldsymbol{\mathcal{X}}^{FC}$, $\boldsymbol{\mathcal{X}}^{DC}$ represent the features respectively extracted from functional connectivity and spatial distance based view, $\parallel$ is the concatenate operation.

\subsection{Domain Generalization}
\begin{figure}[htb]
    \centering
    \includegraphics[width=0.65\columnwidth]{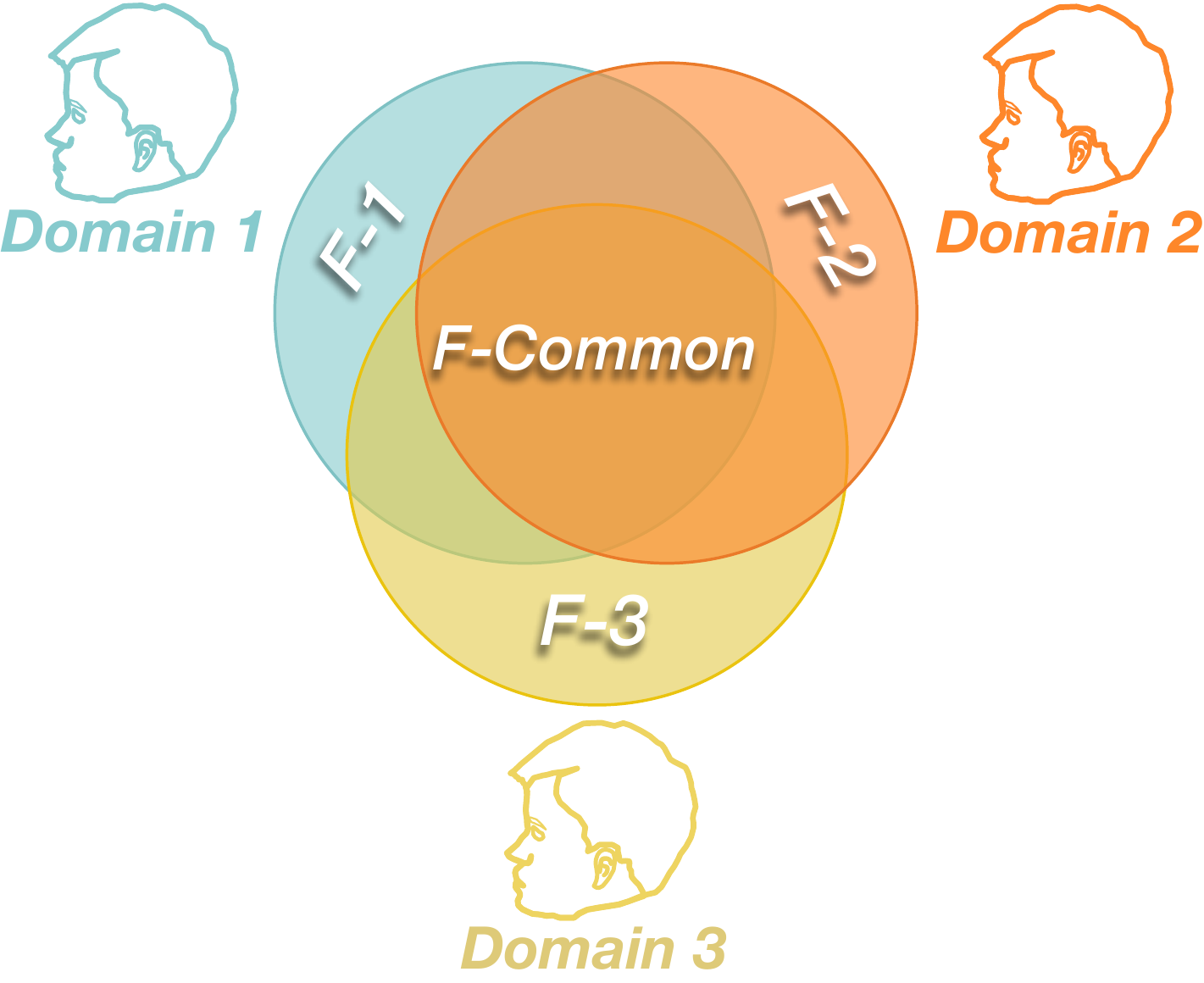}
    \caption{The intuitive idea of the adversarial domain generalization to extract subject-invariant features. Each subject is treated as a specific domain. F-Common means that all subjects have common features for sleep stage classification. F-1, F-2, and F-3 represent some of the subjects' unique features related to sleep stage classification. Domain generalization makes the model unable to distinguish which subject the sample comes from. At the same time, as much as possible to improve the model performance for sleep stage classification. This means that some unique features of subjects are not learned by the model, but some common subject-invariant features (F-Common) related to sleep stage classification are extracted. Therefore, the generalization of the model is improved through domain generalization.}
    \label{DG}
\end{figure}
In order to reduce the influence of individual differences, we exploit an adversarial domain generalization method to enhance the robustness of our model. Figure \ref{DG} presents the intuitive idea of the adversarial domain generalization. Specifically, this method aims to make it impossible to distinguish which source domain the sample data originated from during model training. At the same time, it aims to improve the sleep stage classification performance as much as possible. This means that all subjects' common features (subject-invariant features) related to sleep stage classification are extracted. For example, the model cannot distinguish that the samples of Domain 1 are the data belonging to its own domain, but it can still accurately identify the sleep stages. This presents that the model did not learn personalized features (F-1) belonging to Domain 1, but some common features related to sleep stage classification. In fact, previous studies have presented the advantages of adversarial domain generalization \cite{li2018deep}, and theoretically this method aligns the  marginal distribution of different domains.
Specifically, domain generalization includes three parts: feature extractor $\mathcal{G}_f$, domain classifier $\mathcal{G}_d$ and label predictor $\mathcal{G}_y$. The feature extractor $\mathcal{G}_f$ maps the input data to a domain-invariant feature space,
\begin{equation}
    \mathcal{G}_{f}\left(\boldsymbol{X} ; \theta_{f}\right)= \boldsymbol{\mathcal{X}}
\end{equation}
where $\boldsymbol{X}$ is the input feature matrix, $\theta_{f}$ is the trainable parameter and $\boldsymbol{\mathcal{X}}$ is the transferred feature matrix. 

The transferred features are put into label predictor $\mathcal{G}_y$ and domain classifier $\mathcal{G}_d$ with softmax function:
\begin{equation}
    \hat{y}_{i} = \frac{\exp \left(w_y{\mathcal{X}}_{i}+b_y\right)} {\sum_{i=1}^{N} \exp \left(w_y{\mathcal{X}}_{i}+b_y\right)}
\end{equation}
\begin{equation}
    \hat{d}_{i} = \frac{\exp \left(w_d{\mathcal{X}}_{i}+b_d\right)} {\sum_{i=1}^{N} \exp \left(w_d{\mathcal{X}}_{i}+b_d\right)}
\end{equation}
where $\mathcal{X}_{i}$ denotes the transferred features of sample $i$. $\hat{y}_{i}$ and $\hat{d}_{i}$ are the predicted results of $\mathcal{G}_y$ and $\mathcal{G}_d$, respectively. Both of the $\mathcal{G}_y$ and $\mathcal{G}_d$ are multi-class classifier, we employ the cross entropy as the loss function: 
\begin{equation}
\mathcal{L}_{y}=-\frac{1}{L} \sum_{i=1}^{L}\sum_{r=1}^{R_y} y_{i, r} \log \hat{y}_{i, r}
\end{equation}
\begin{equation}
\mathcal{L}_{d}=-\frac{1}{L} \sum_{i=1}^{L}\sum_{r=1}^{R_d} d_{i, r} \log \hat{d}_{i, r}
\end{equation}
where $\mathcal{L}_{y}$ is the cross entropy loss function of the multi-classification task, $L$ denotes the number of samples, $R_y$ and $R_d$ denote the number of classes and the number of domains, respectively. $y$ is the true label and $\hat{y}$ is the value predicted by the model. $d$ is the true domain and $\hat{d}$ is the value predicted by the model.

Besides, a special layer called Gradient Reversal Layer (GRL) is implemented between feature extractor $\mathcal{G}_f$ and domain classifier $\mathcal{G}_d$ to form an adversarial relationship \cite{ganin2015unsupervised}.
Compared with other methods that usually require training classifier and discriminator in separate steps, GRL can integrate feature learning and domain generalization in a unified framework and execute backpropagation algorithms. 
The optimization process is defined as:
\begin{equation}
    \begin{array}{l} 
        \left( {{{\hat \theta }_f},{{\hat \theta }_y}} \right) = arg\mathop {\min }\limits_{{\theta _f},{\theta _y}} \mathcal{L}\left( {{\theta _f},{\theta _y},{{\hat \theta }_d}} \right) \\ 
        \left( {{{\hat \theta }_d}} \right) = arg\mathop {\max }\limits_{{\theta _d}} \mathcal{L}\left( {{{\hat \theta }_f},{{\hat \theta }_y},{\theta _d}} \right) \\ 
    \end{array}
\end{equation}
where $\theta _d, \theta _y$ are the parameters to minimize the loss of $\mathcal{G}_d$ and $ \mathcal{G}_y$, respectively. $\theta _f$ is the parameters of $\mathcal{G}_f$ to minimize the loss of $\mathcal{G}_y$ and maximize the loss of $\mathcal{G}_d$ at the same time. The aims of feature extractor $\mathcal{G}_f$ and domain classifier $\mathcal{G}_d$ are exact opposite. The feature extractor $\mathcal{G}_f$ aims to make the domain classifier $\mathcal{G}_d$ can't classify the right domain and the domain classifier $\mathcal{G}_d$ aims to correctly classify the domain that the data comes from.

The whole loss function of the domain generalization is defined as:
\begin{equation}
  \mathcal{L}_{\text {DG}} =
    -\frac{1}{L} \sum_{i=1}^{L}\sum_{r=1}^{R_y} y_{i, r} \log \hat{y}_{i, r}
    + \beta \frac{1}{L} \sum_{i=1}^{L}\sum_{r=1}^{R_d} d_{i, r} \log \hat{d}_{i, r}
\end{equation}
By optimizing the loss function, the feature extractor $\mathcal{G}_f$ can achieve the goal of finding the domain-invariant feature space.

\section{Experiments and Discussions}

\subsection{Dataset and Experiment Settings}
Two publicly available datasets are employed in our experiments: 1) \textbf{ISRUC-S3 dataset} \cite{khalighi2016isruc} contains 10 healthy subjects (9 male and 1 female). Each recording contains 6 EEG channels, 2 EOG channels, 3 EMG channels, and 1 ECG channel. In addition, the experts classify these PSG recordings into five sleep stages according to AASM standard\cite{berry2012rules}.
2) \textbf{MASS-SS3 dataset} \cite{o2014montreal} contains 62 healthy subjects (28 male and 34 female). Each recording contains 20 EEG channels, 2 EOG channels, 3 EMG channels, and 1 ECG channel. 
\begin{table*}[htb]
\centering
\caption{The performance comparison of the state-of-the-art approaches on the ISRUC-S3 dataset}
\begin{threeparttable}[l]
\setlength{\tabcolsep}{3.9mm}{
\begin{tabular}{@{}lccccccccc@{}}
\toprule
\multicolumn{1}{c}{\multirow{2}{*}{}} & \multirow{2}{*}{Method} & \multicolumn{3}{c}{Overall results} & \multicolumn{5}{c}{F1-score for each class} \\
\cmidrule{3-10} 
                                &               & Accuracy            & F1-score          & Kappa            & Wake    & N1     & N2     & N3     & REM    \\ \cmidrule{1-10}
Alickovic et al.\cite{alickovic2018ensemble}            & SVM           &0.733	&0.721	&0.657	&0.868	&0.523	&0.699	&0.786	&0.731  \\
Memar et al.\cite{memar2017novel}            & RF            &0.729	&0.708	&0.648	&0.858	&0.473	&0.704	&0.809	&0.699   \\
Dong et al.\cite{dong2017mixed}            & MLP+LSTM       &0.779	&0.758	&0.713	&0.860	&0.469	&0.760	&0.875	&0.828   \\
Supratak et al.\cite{supratak2017deepsleepnet} & CNN+BiLSTM    &0.788	&0.779	&0.730	&\underline{0.887}	&\textbf{0.602}	&0.746	&0.858	&0.802   \\
Chambon et al.\cite{chambon2018deep}          & CNN           &0.781	&0.768	&0.720	&0.870	&0.550	&0.760	&0.851	&0.809  \\
Phan et al.\cite{phan2019seqsleepnet}      & ARNN+RNN      &0.789	&0.763	&0.725	&0.836	&0.439	&\underline{0.793}	&\underline{0.879}	&\textbf{0.867}  \\
Jia et al. \cite{jia2020graphsleepnet} &STGCN            &\underline{0.799} &\underline{0.787} &\underline{0.741} &0.878 &0.574  &0.776  &0.864  &0.841 \\
\cmidrule{1-10}
\textbf{proposed model}                   & \textbf{MSTGCN}         & \textbf{0.821}      & \textbf{0.808}             & \textbf{0.769}   & \textbf{0.894}   & \underline{0.596}  & \textbf{0.806}  & \textbf{0.890}  & \underline{0.856}  \\ 
\bottomrule
\end{tabular}
}
    \begin{tablenotes}
     \item[*] The bold result is the best result and the underlined result is the second best result.
   \end{tablenotes}
\end{threeparttable}
\label{ISRUCresult}
\end{table*}

\begin{table*}[htb]
\centering
\caption{The performance comparison of the state-of-the-art approaches on the MASS-SS3 dataset}
\begin{threeparttable}[l]
\setlength{\tabcolsep}{3.9mm}{
\begin{tabular}{@{}lccccccccc@{}}
\toprule
\multicolumn{1}{c}{\multirow{2}{*}{}} & \multirow{2}{*}{Method} & \multicolumn{3}{c}{Overall results} & \multicolumn{5}{c}{F1-score for each class} \\ \cmidrule{3-10} 
                                &               & Accuracy            & F1-score          & Kappa            & Wake    & N1     & N2     & N3     & REM    \\ \cmidrule{1-10}
Alickovic et al.\cite{alickovic2018ensemble}            & SVM           & 0.779 &0.688 &0.659 &0.801 &0.339 &0.843 &0.645 &0.813  \\
Memar et al.\cite{memar2017novel}            & RF            & 0.800 &0.726 &0.697 &0.863 &0.379 &0.858 &0.784 &0.749  \\
Dong et al.\cite{dong2017mixed}            & MLP+LSTM       & 0.859               & 0.805    & -                & 0.846   & 0.563  & 0.907  & 0.848  & 0.861  \\
Supratak et al.\cite{supratak2017deepsleepnet} & CNN+BiLSTM    & 0.862               & 0.817             & 0.800            & 0.873   & 0.598  & 0.903  & 0.815  & 0.893  \\
Chambon et al.\cite{chambon2018deep}          & CNN           & 0.739               & 0.673             & 0.640            & 0.730   & 0.294  & 0.812  & 0.765  & 0.764  \\
Phan et al.\cite{phan2019seqsleepnet}      & ARNN+RNN      & 0.871               & 0.833             & 0.815            & -       & -      & -      & -      & -      \\
Jia et al. \cite{jia2020graphsleepnet} &STGCN  & \underline{0.889}  & \underline{0.841}  & \underline{0.834} &\textbf{0.913}  & \underline{0.603}  & \underline{0.921}  & \underline{0.851}  & \underline{0.919} \\
\cmidrule{1-10}
\textbf{proposed model}  & \textbf{ MSTGCN}  & \textbf{0.895}  & \textbf{0.854}  & \textbf{0.843}  & \underline{0.911}   & \textbf{0.645}  & \textbf{0.924}  & \textbf{0.866}  & \textbf{0.924}  \\ 
\bottomrule
\end{tabular}}
    \begin{tablenotes}
     \item[*] The bold result is the best result and the underlined result is the second best result.
   \end{tablenotes}
\end{threeparttable}
\label{MASSresult}
\end{table*}

We compare our MSTGCN with 7 baselines, which are described in detail in Supplementary Material S.3.
For a fair comparison, we employ the same experimental settings for all models. Specifically, we employ 10-fold cross-validation and 31-fold cross-validation to evaluate the performance of all models on ISRUC-S3 dataset and MASS-SS3 dataset, respectively.  In addition, we adopt the subject-independent strategy for cross-validation. 
We implement the proposed model using TensorFlow. In addition, the code is released on Github$\footnote{https://github.com/ziyujia/MSTGCN}$.


\subsection{Comparison with the State-of-the-Art Methods}

We compare the proposed model with the other baseline models for sleep stage classification on the ISRUC-S3 and MASS-SS3 as presented in Table \ref{ISRUCresult} and Table \ref{MASSresult}. The results present that our proposed model outperforms the baseline methods on multiple overall metrics (overall Accuracy, F1-score, and Kappa) for ISRUC-S3 and MASS-SS3.
Specifically, the traditional machine learning methods (SVM and RF) cannot learn the complex spatial or temporal features well. However, existing deep learning models such as CNN and RNN \cite{dong2017mixed,supratak2017deepsleepnet,chambon2018deep,phan2019seqsleepnet} can directly extract the spatial or temporal features. Therefore, their performance is better than the traditional machine learning methods.

Although CNN and RNN achieve high accuracy, their limitation is that the model's input must be grid data ignoring the connection among brain regions. Due to brain regions are in non-Euclidean space, graph is the most appropriate data structure to indicate the connections. Therefore, the proposed model and ST-GCN can often achieve optimal or suboptimal overall results, especially on the MASS-SS3 dataset. In addition, the proposed model extracts both spatial and temporal features based on multi-view brain graphs and integrates domain generalization to learn subject-invariant features. Hence, the proposed model achieves the state-of-the-art performance.


For different sleep stages, MSTGCN can accurately identify most of the corresponding stages. Specifically, in the ISRUC-S3 dataset, the classification accuracy of Wake and N3 stages is the highest. In the MASS-SS3 dataset, the classification accuracy of the REM and N2 stages is the highest. However, the classification performance of the N1 stage does not meet expectations on the two datasets, like other baseline models.  It may be because the N1 stage is a transitional period between the Wake stage and the N2 stage, and the sample number of N1 stage is relatively small.
Therefore, as Figure S.2 in Supplementary Material shows, N1 stage is mistakenly divided into other sleep stages, such as Wake stage and N2 stage. 
Nevertheless, the classification performance of MSTGCN for the N1 stage is still higher than most baseline models. Table \ref{MASSresult} presents that MSTGCN has the highest F1-score for N1 stage on the MASS-SS3 dataset, which is 4\% higher than the sub-optimal result.


\subsection{Experimental Analysis and Discussion}
\subsubsection{Ablation Experiment}
To validate the effect of each module in our model, we design some variant models. First, we use the spatial graph convolution with spatial distance brain graph as the basic model to gradually stack the remaining modules to form a whole branch. Then, we add another whole ST-GCN branch with functional connectivity brain graph to form a multi-view ST-GCN. Finally, we integrate the domain generalization method to form the proposed model. The specific process is described as follows:
\begin{itemize}
\item{\emph{variant a (Spatial Graph Convolution (Base Model)):}} We utilize a spatial graph convolution network with spatial distance brain graph as the base model.

\item{\emph{variant b (+ Temporal Convolution):}} We add temporal convolution to form a spatial-temporal graph convolution network.

\item{\emph{variant c (+ Attention Mechanism):}} We add attention mechanism both on spatial and temporal dimension.

\item{\emph{variant d (+ Multi-view Fusion (Add Another View)):}} We add another whole ST-GCN branch based on the functional connectivity brain graph to form a multi-view ST-GCN.

\item{\emph{variant e (+ Domain Generalization):}} A multi-view ST-GCN with domain generalization (our MSTGCN).
\end{itemize}

\begin{figure}[htb]
    \centering
    \includegraphics[height=38mm, trim={0 14.8 0 11.5}, clip]{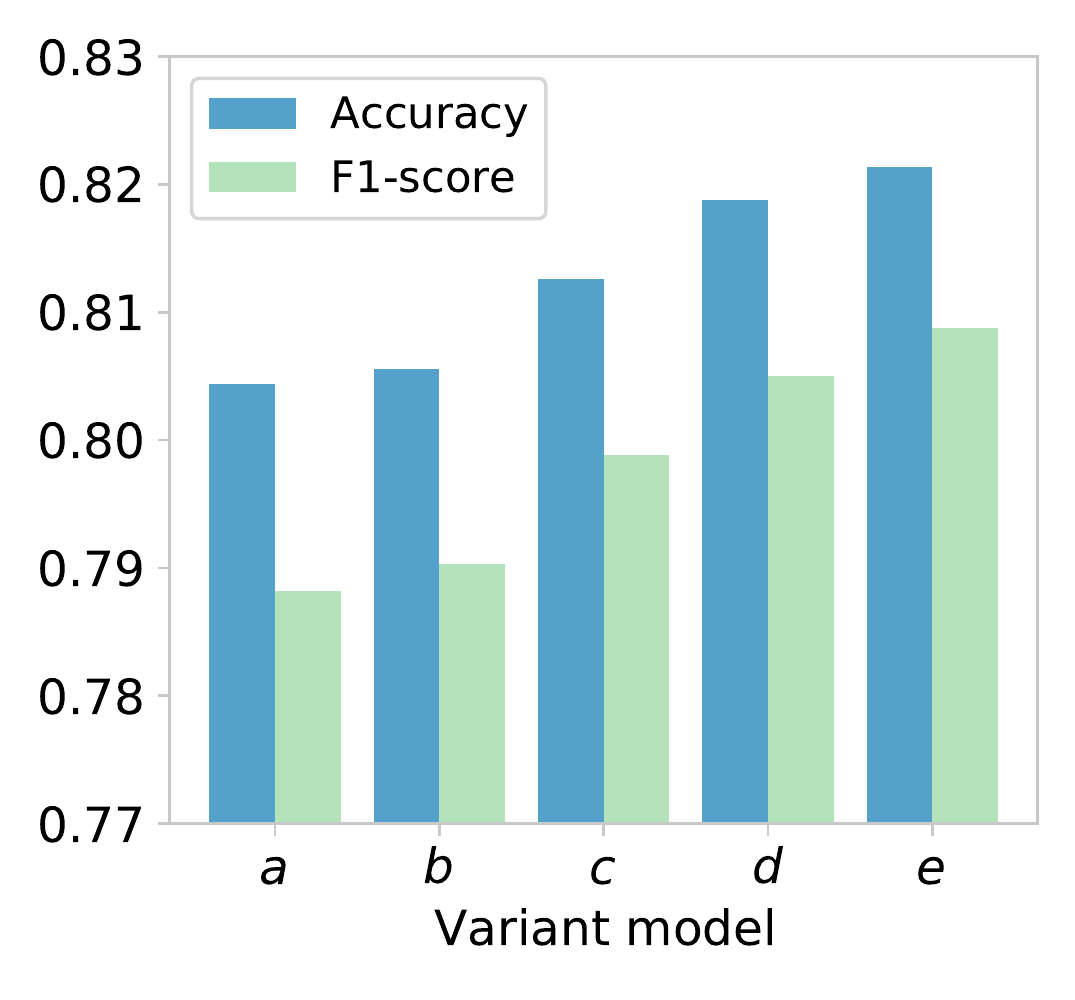}
    \caption{Comparison of the designed variant models to verify the effectiveness of different modules in MSTGCN.}
    \label{Ablation}
\end{figure}

Figure \ref{Ablation} presents that the key modules in our model are effective for sleep stage classification, especially variant \textit{c}, variant \textit{d}, and variant \textit{e}. Specifically, the attention mechanism helps to capture valuable spatial-temporal features to improve the classification performance of our model. The designed multi-view on brain provides complementary information for sleep stage classification. In addition, domain generalization is integrated into the multi-view ST-GCN to extract subject-invariant features, which helps to improve the model generalization. In summary, the ablation experiment presents the effectiveness of each module in our model.

\subsubsection{Adaptive Functional Connectivity Graph}
To further investigate the effectiveness of the adaptive functional connectivity graph learning, we design five fixed functional connectivity graphs to compare with it. These graphs are defined as different adjacency matrices. The last three graphs are constructed by functional connectivity methods commonly found in neuroscience.
\begin{itemize}
\item Fully Connected Adjacency Matrix: A matrix whose elements are all 1. It represents that there are all connections among all nodes and each node also has self-connection in the graph.  
\item ${K}$-Nearest Neighbor (KNN) Adjacency Matrix \cite{jiang2013graph}: A matrix, which represents a $k$-nearest neighbor graph. That is, each node has $k$ neighbor nodes. 
\item Pearson Correlation Coefficient (PCC) Adjacency Matrix \cite{pearson1903laws}: A matrix generated by the pearson correlation coefficient between each pair of nodes.  
\item Phase Locking Value (PLV) Adjacency Matrix \cite{aydore2013note}: A matrix generated by the PLV method between each pair of nodes.
\item Mutual Information (MI) Adjacency Matrix \cite{danon2005comparing}: A matrix generated by measuring the mutual dependence between each pair of nodes.
\end{itemize}

\begin{figure}[htb]
    \centering
    \includegraphics[height=39mm, trim={0 11.5 0 11.8}, clip]{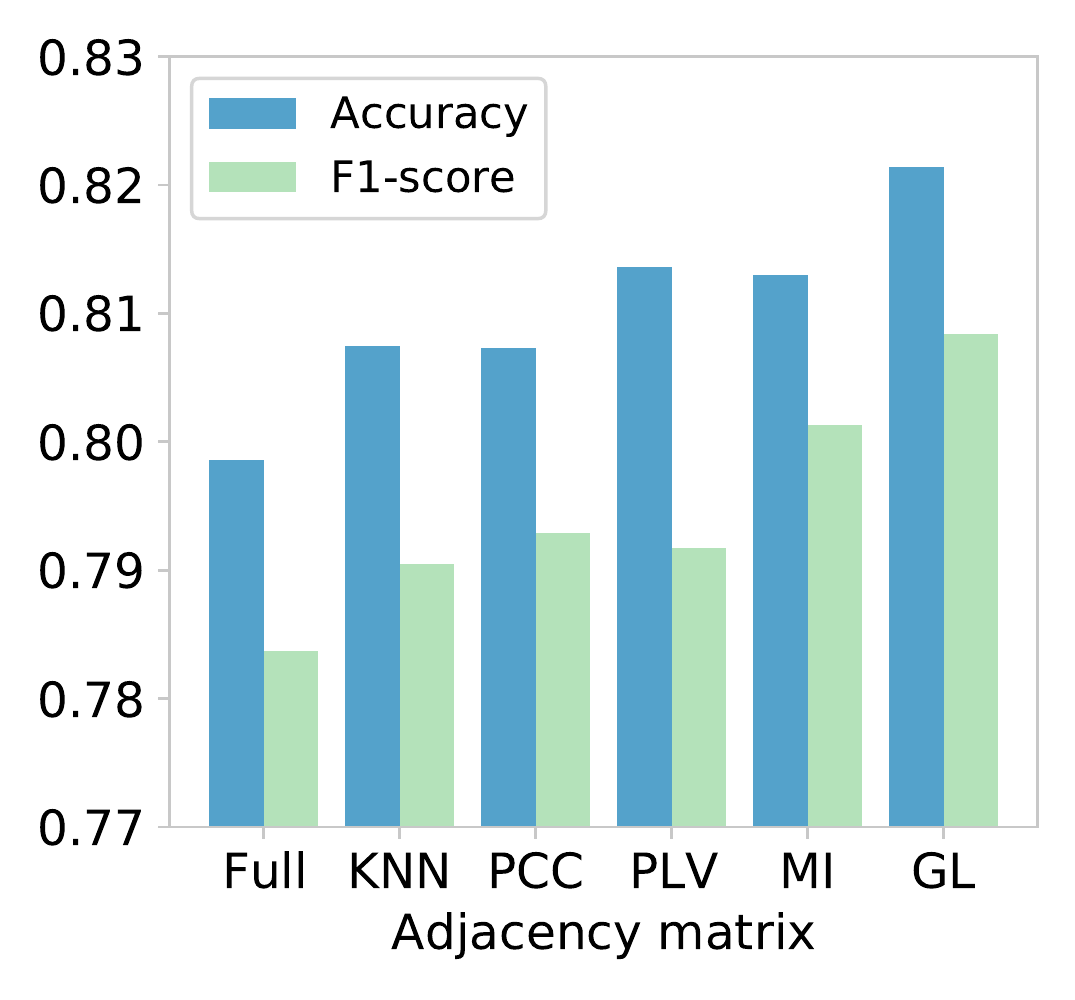}
    \caption{Comparison of different adjacency matrices. GL: the proposed Graph Learning approach for brain functional connectivity. Full: Fully Connected Adjacency Matrix; KNN: ${K}$-Nearest Neighbor Adjacency Matrix; PCC: Pearson Correlation Coefficient Adjacency Matrix; PLV: Phase Locking Value Adjacency Matrix; MI: Mutual Information Adjacency Matrix. }
    \label{adjacency matrices}
\end{figure}
Figure \ref{adjacency matrices} illustrates that the adaptive (learned) adjacency matrix achieves the highest accuracy for sleep stage classification. In addition, the adjacency matrix combined with prior neuroscience knowledge also achieves a suboptimal effect, such as the PCC, PLV, and MI adjacency matrix. The fully connected adjacency matrix does not work well because the brain network is not a fully connected graph. In general, the adjacency matrix can significantly affect the classification performance. The proposed adaptive functional connectivity graph for classification tasks is superior to the fixed functional connectivity graphs.

\begin{figure*}[htb]
\centering 
\subfigure[N1 Stage]{
\includegraphics[height=28mm, trim={13 17 64.5 14}, clip]{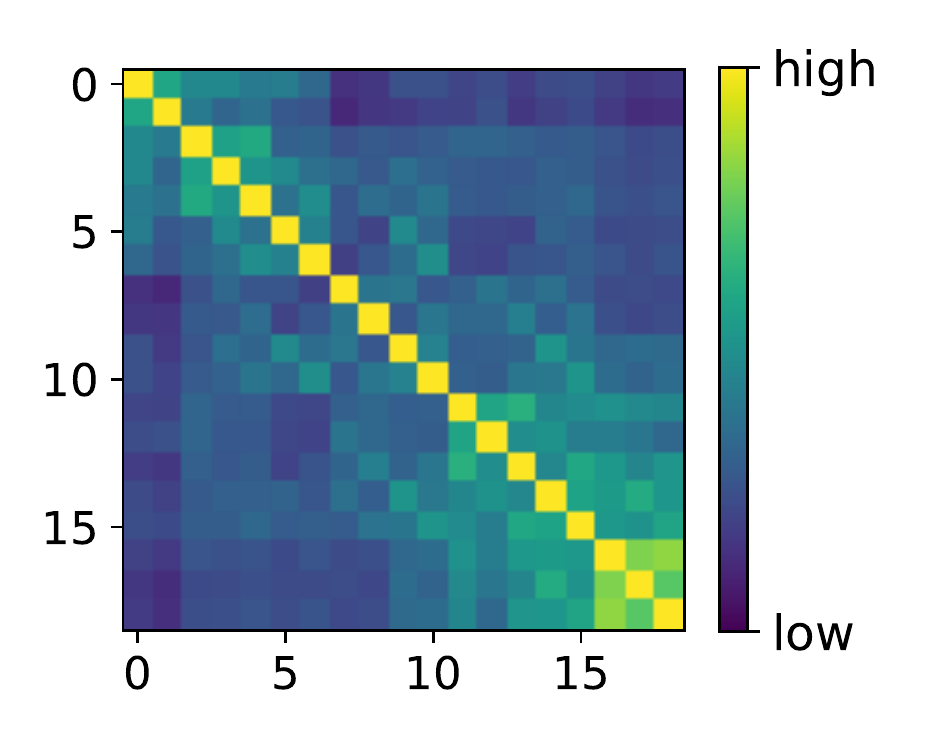}}
\subfigure[N2 Stage]{
\includegraphics[height=28mm, trim={13 17 64.5 14}, clip]{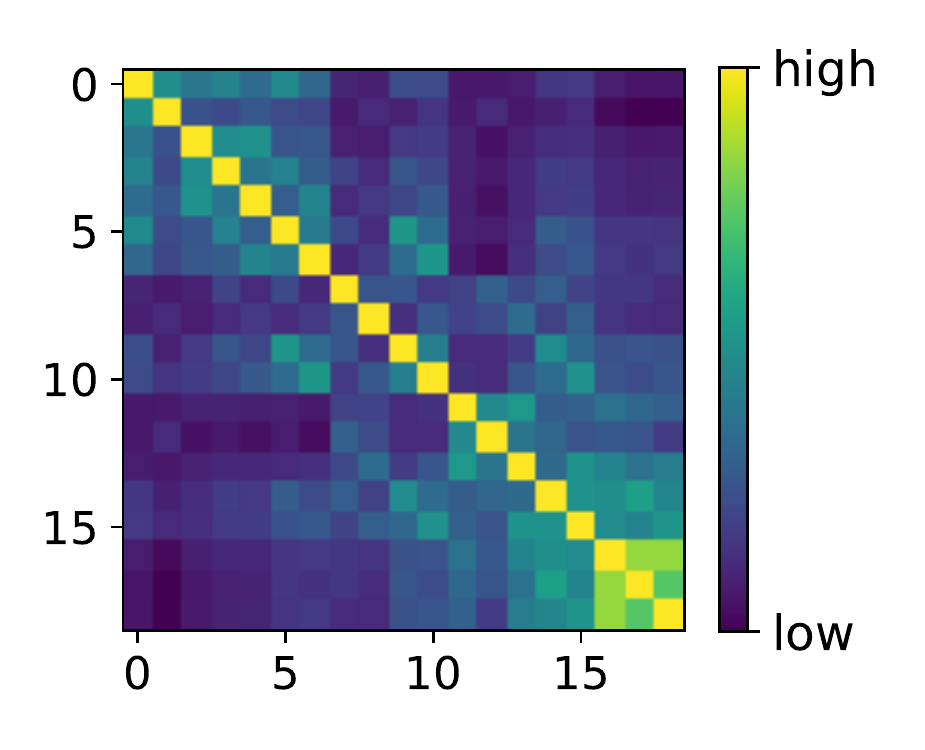}}
\subfigure[N3 Stage]{
\includegraphics[height=28mm, trim={13 17 64.5 14}, clip]{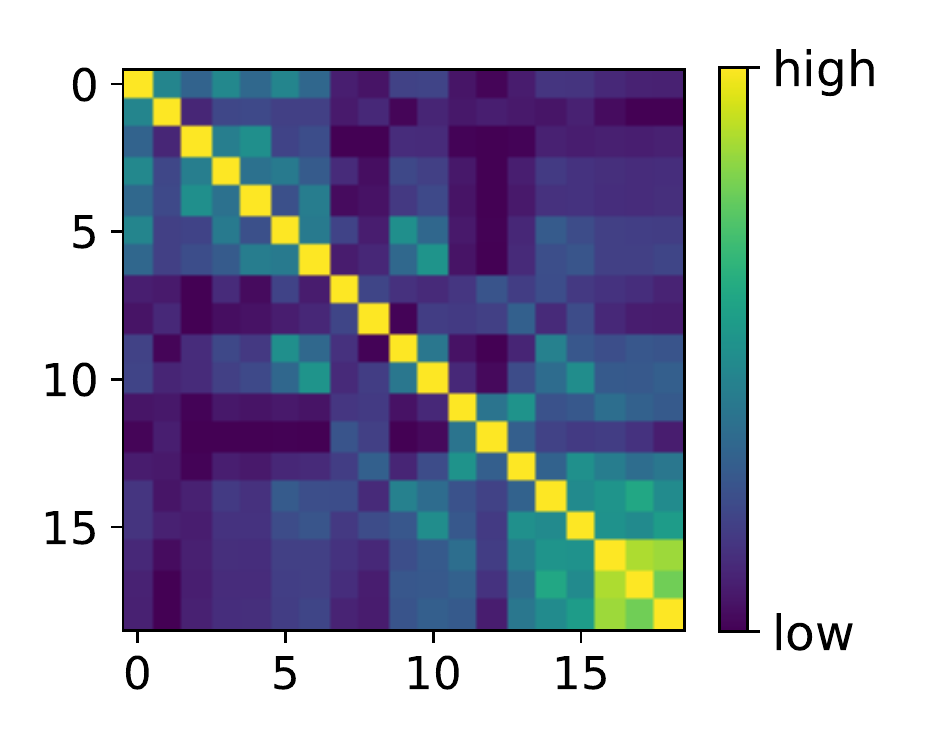}}
\subfigure[Wake Stage]{
\includegraphics[height=28mm, trim={13 17 64.5 14}, clip]{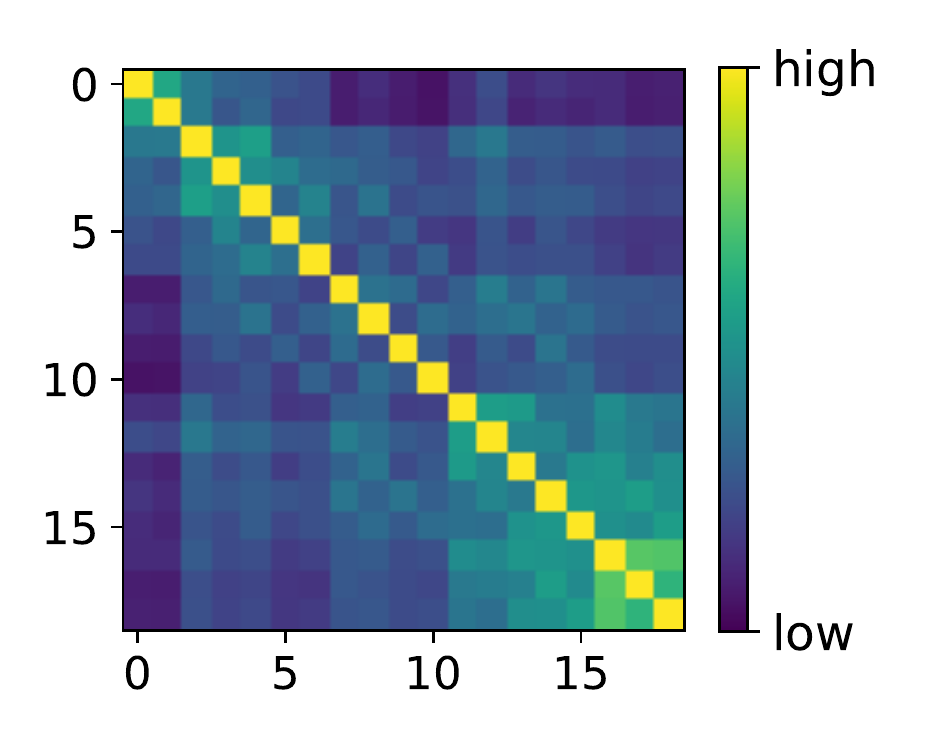}}
\subfigure[REM Stage]{
\includegraphics[height=28mm, trim={13 17 14.5 14}, clip]{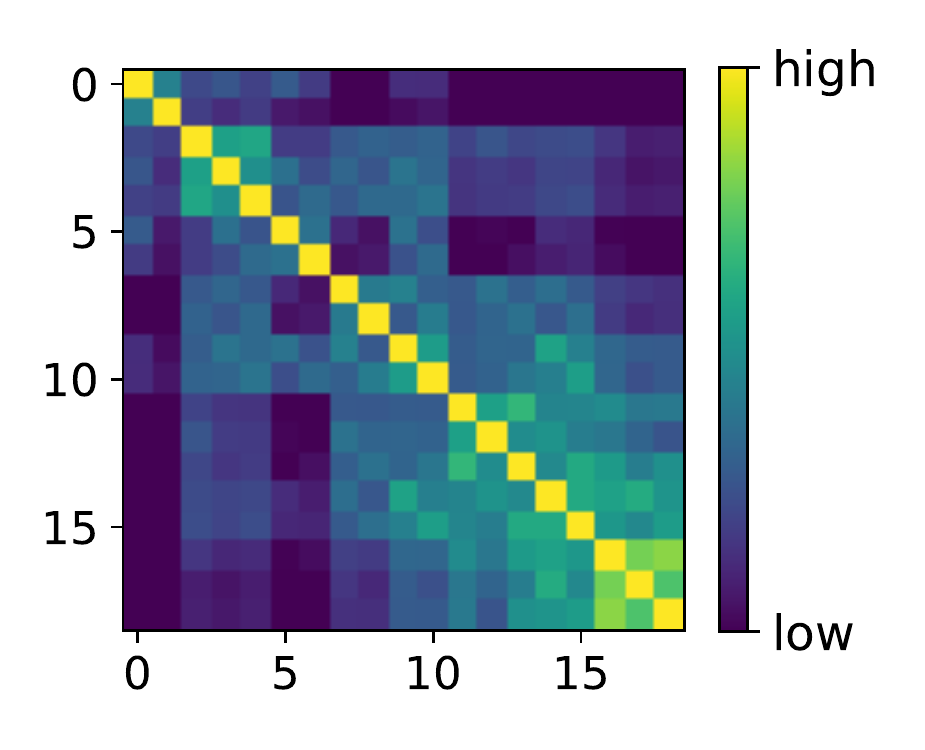}}
\caption{The learned adjacency matrix visualization of five sleep stages (N1 Stage, N2 Stage, N3 Stage, Wake Stage, and REM Stage).}
\label{adjacency matrix visualization}
\end{figure*}
To present the interpretability of the adaptive functional connectivity graph, we visualize the brain adjacency matrices obtained by adaptive learning for different sleep stages. These matrices reflect the brain functional connectivity in different sleep stages as illustrated in Figure \ref{adjacency matrix visualization}. Specifically, there are more functional connectivity in the Wake stage and N1 stage. On the contrary, the functional connectivity of the N3 stage is the least. These findings are consistent with existing neuroscience research \cite{spoormaker2010development, larson2011modulation}. N3 stage is a typical deep sleep period, and the brain is usually in an inactive stage. In contrast, the N1 stage is a light sleep period, and the brain is relatively active. Therefore, the functional connectivity of the brain in the N1 stage is relatively complicated.

\begin{figure}[htb]
\centering
\includegraphics[height=39mm, trim={0 16 0 15.5}, clip]{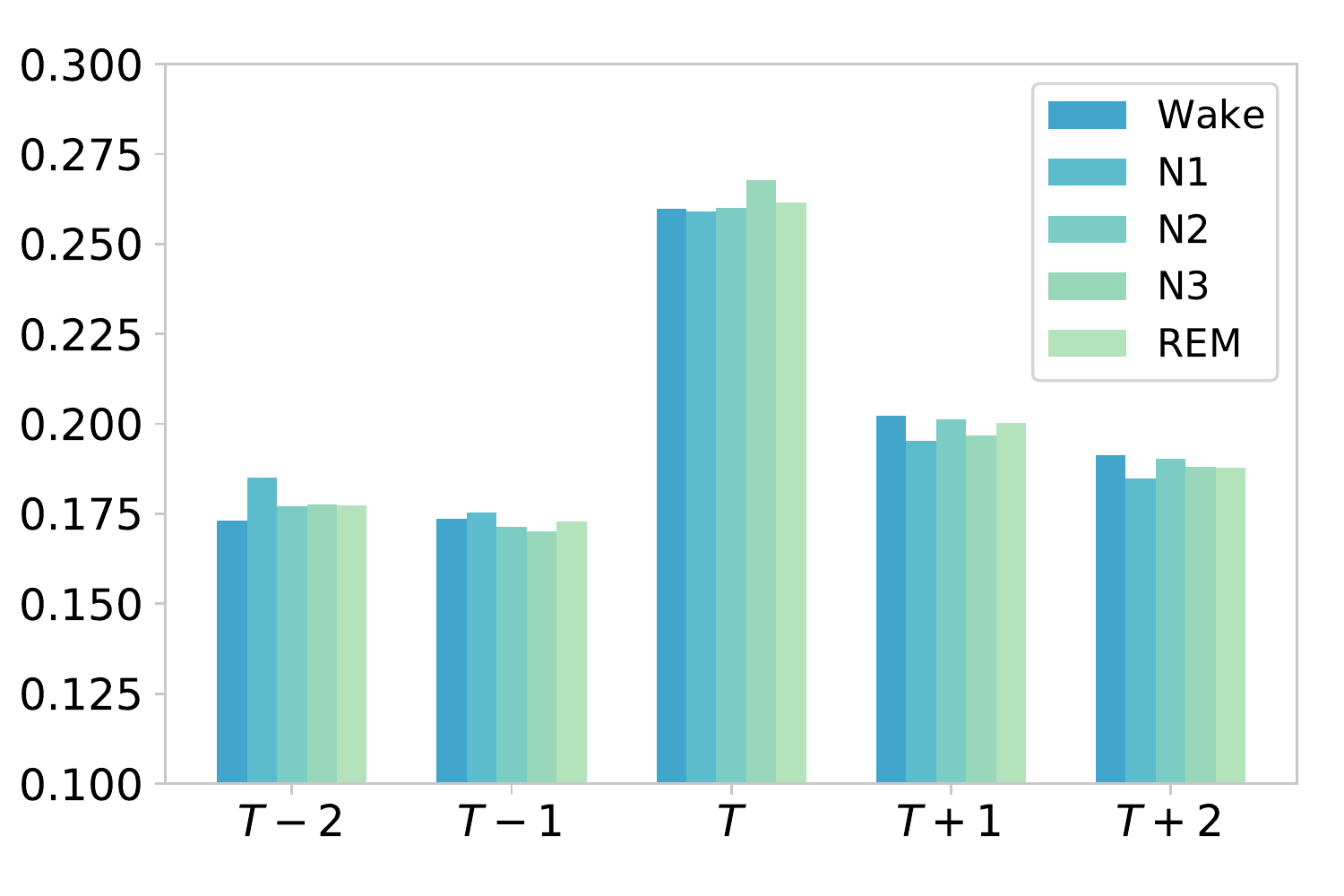}
\caption{Temporal attention visualization. The current sleep stage $T$ always keeps the most attention weights. The adjacent sleep stages keep some attention weights for this classification task.} 
\label{TATT}
\end{figure}

\subsubsection{Attention Mechanism} To explore the interpretability of the attention mechanism, first we visualize the learned weight of temporal attention mechanism to indicate the importance of different sleep epochs for classification. The higher the weight, the higher the degree of attention. Figure \ref{TATT} illustrates that the weight of the current sleep stage $T$ is the largest. Previous and following sleep epochs received similar but lower attention. That is, this stage has received the most attention, which is consistent with the AASM sleep standard \cite{berry2012rules}. In fact, sleep experts mainly judge the current sleep stage type based on the characteristics of the current sleep state and appropriately refer to the adjacent sleep state. Therefore, the temporal attention mechanism has learned expert knowledge to a certain extent.

\begin{figure}[htb]
\centering
\includegraphics[height=37mm, trim={0 16 0 14}, clip]{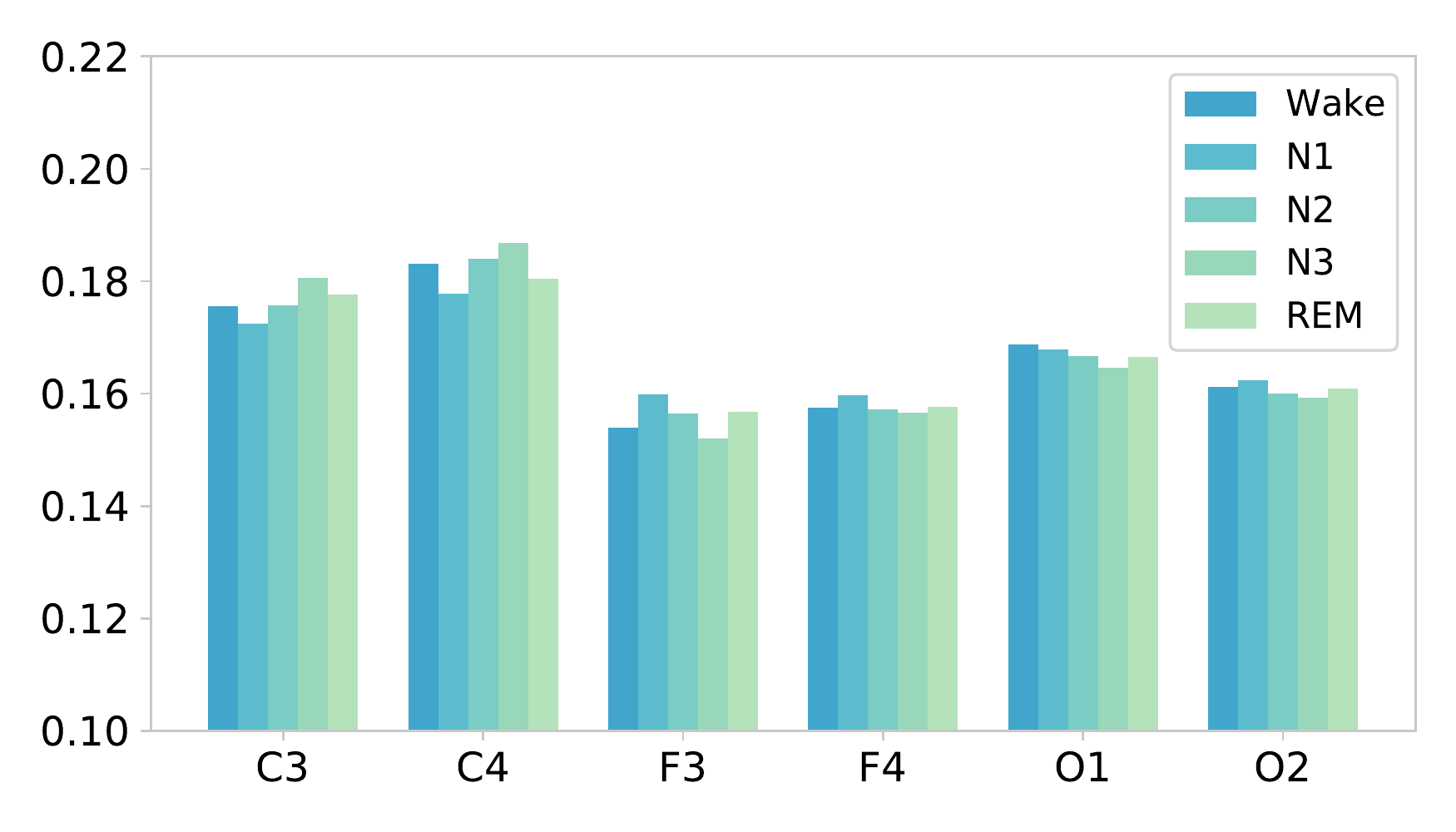} 
\caption{Spatial attention visualization to present the contribution of various EEG channels for sleep stage classification. The attention weights of C3 channel and C4 channel is always the highest for different sleep stages. } 
\label{SATT}
\end{figure}

In addition, we also visualize the learned weight of spatial attention mechanism for EEG channels. Figure \ref{SATT} illustrates that our model pays different attention to EEG channels in different sleep stages, which may caused by the EEG patterns of different sleep stages are different. The attention weights of F3 and F4 are always the lowest. In contrast, the attention weights of C3 and C4 have always been the highest for different sleep stages.  The results indicate that C3 and C4 may be the most informative EEG channels for sleep stage classification. Generally, the C3 and C4 channels are located in the middle of the scalp, which may have richer EEG information and be less affected by external factors.

\section{Conclusion}

In this paper, we propose a novel deep graph neural network MSTGCN for sleep stage classification. In MSTGCN, we propose effective approaches in modeling the dynamics of sleep data along both the spatial and temporal dimensions, as well as considering the subject differences in sleep data.  Specifically, we design different brain views based on the functional connectivity and physical distance proximity of the brain. The complementarity of different views provides rich spatial topology information. We develop a spatial-temporal graph convolution with attention mechanism to simultaneously capture the most relevant spatial-temporal features for sleep stage classification. Moreover, to extract subject-invariant sleep features, we integrate domain generalization and spatial-temporal graph convolutional networks into a unified framework. Experiments on two public sleep datasets demonstrate MSTGCN achieves the state-of-the-art performance. Finally, our proposed approach provides a general-framework for multivariate physiological time series.

\ifCLASSOPTIONcaptionsoff
  \newpage
\fi



\bibliographystyle{IEEEtran}
\bibliography{ref}
\clearpage
\includepdfmerge{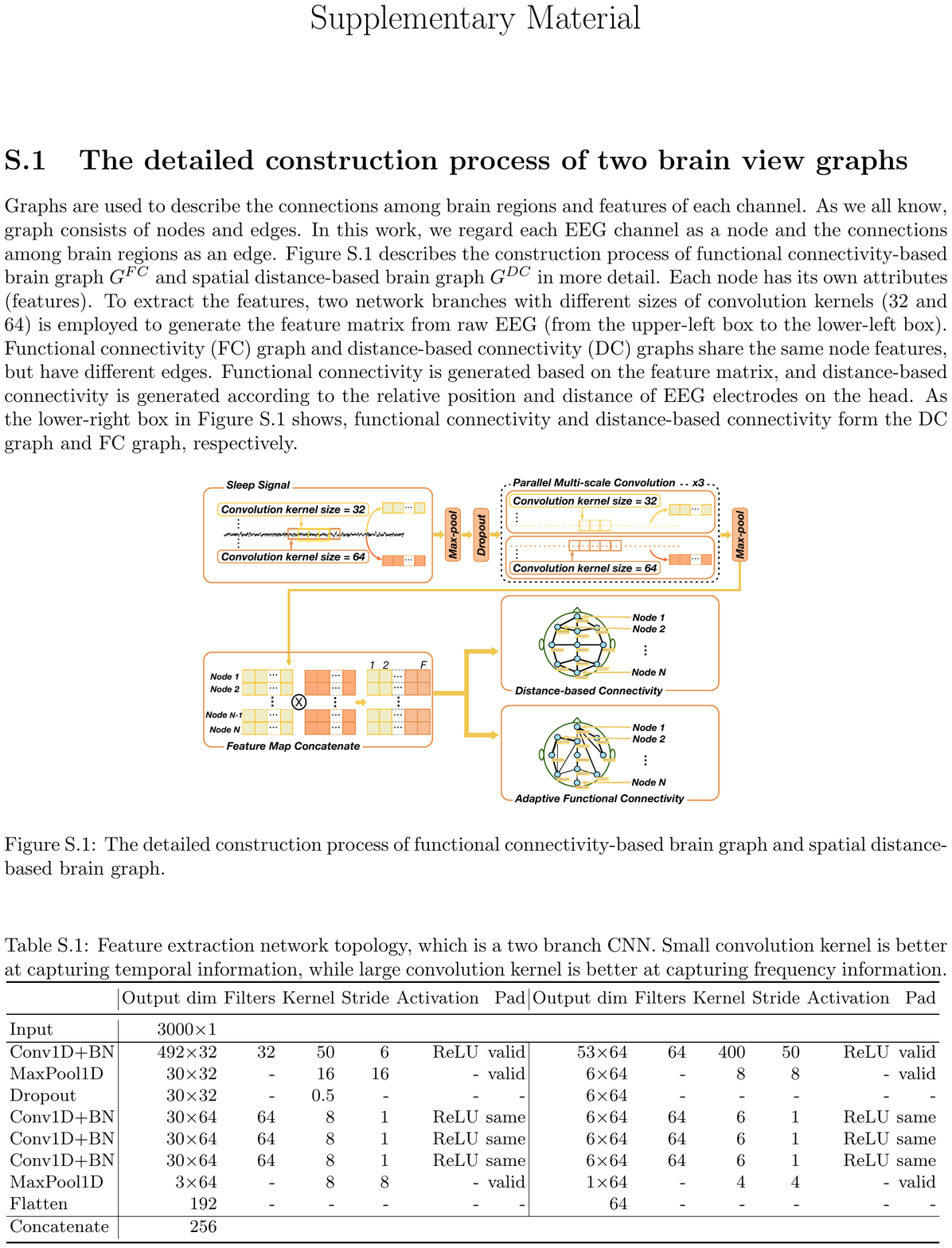,1-4}


\end{document}


\maketitle

\section{The detailed construction process of two brain view graphs}


Graphs are used to describe the connections among brain regions and features of each channel. As we all know, graph consists of nodes and edges. In this work, we regard each EEG channel as a node and the connections among brain regions as an edge.
Figure \ref{FC_DC_Graph} describes the construction process of functional connectivity-based brain graph $G ^ {FC}$ and spatial distance-based brain graph $G ^ {DC}$ in more detail.
Each node has its own attributes (features). To extract the features, two network branches with different sizes of convolution kernels (32 and 64) is employed to generate the feature matrix from raw EEG (from the upper-left box to the lower-left box). 
Functional connectivity (FC) graph and distance-based connectivity (DC) graphs share the same node features, but have different edges. Functional connectivity is generated based on the feature matrix, and distance-based connectivity is generated according to the relative position and distance of EEG electrodes on the head. As the lower-right box in Figure \ref{FC_DC_Graph} shows, functional connectivity and distance-based connectivity form the DC graph and FC graph, respectively.

\begin{figure}[htb]
    \centering
    \includegraphics[width=0.58\textwidth]{journal_feature_210627_1.pdf}
    \caption{The detailed construction process of functional connectivity-based brain graph and spatial distance-based brain graph.}
    \label{FC_DC_Graph}
\end{figure}

\begin{table}[htb]
\centering
\caption{Feature extraction network topology, which is a two branch CNN. Small convolution kernel is better at capturing temporal information, while large convolution kernel is better at capturing frequency information.}
\setlength{\tabcolsep}{0.6mm}{
\small
\begin{tabular}{l|rrrrrrrrrrrr}
\toprule
            & Output dim & Filters & Kernel & Stride & Activation & \multicolumn{1}{r|}{Pad}   & Output dim & Filters & Kernel & Stride & Activation & Pad   \\ \midrule
Input       & 3000×1     &         &        &        &            &                            &            &         &        &        &            &       \\ \hline
Conv1D+BN   & 492×32     & 32      & 50     & 6      & ReLU       & \multicolumn{1}{r|}{valid} & 53×64      & 64      & 400    & 50     & ReLU       & valid \\
MaxPool1D   & 30×32      & -       & 16     & 16     & -          & \multicolumn{1}{r|}{valid} & 6×64       & -       & 8      & 8      & -          & valid \\
Dropout     & 30×32      & -       & 0.5    & -      & -          & \multicolumn{1}{r|}{-}     & 6×64       & -       & -      & -      & -          & -     \\
Conv1D+BN   & 30×64      & 64      & 8      & 1      & ReLU       & \multicolumn{1}{r|}{same}  & 6×64       & 64      & 6      & 1      & ReLU       & same  \\
Conv1D+BN   & 30×64      & 64      & 8      & 1      & ReLU       & \multicolumn{1}{r|}{same}  & 6×64       & 64      & 6      & 1      & ReLU       & same  \\
Conv1D+BN   & 30×64      & 64      & 8      & 1      & ReLU       & \multicolumn{1}{r|}{same}  & 6×64       & 64      & 6      & 1      & ReLU       & same  \\
MaxPool1D   & 3×64       & -       & 8      & 8      & -          & \multicolumn{1}{r|}{valid} & 1×64       & -       & 4      & 4      & -          & valid \\
Flatten     & 192        & -       & -      & -      & -          & \multicolumn{1}{r|}{-}     & 64         & -       & -      & -      & -          & -     \\ \hline
Concatenate & 256        &         &        &        &            &                            &            &         &        &        &            &      \\
\bottomrule
\end{tabular}
}
\end{table}

\newpage



\section{Dataset statistics}

\begin{table}[htb]
\centering
\caption{Number of 30s epochs for each sleep stage on ISRUC-S3 dataset}
\setlength{\tabcolsep}{1.8mm}{
\begin{tabular}{ccccccc}
\toprule
          & Wake      & N1    & N2     & N3     & REM    & Total \\ \midrule
Sample Number    & 1651   & 1215  & 2609  & 2014   & 1060  & 8549 \\
Proportion & 19.3\% & 14.2\% & 30.5\% & 23.6\% & 12.4\% & 100\% \\ 
\bottomrule
\end{tabular}
}
\label{ISRUC}
\end{table}

\begin{table}[htb]
\centering
\caption{Number of 30s epochs for each sleep stage on MASS-SS3 dataset}
\setlength{\tabcolsep}{1.8mm}{
\begin{tabular}{ccccccc}
\toprule
           & Wake      & N1    & N2     & N3     & REM    & Total \\ \midrule
Sample  Number   & 6277   & 4819  & 29749  & 7651   & 10560  & 59056 \\
Proportion & 10.6\% & 8.2\% & 50.4\% & 13.0\% & 17.9\% & 100\% \\ 
\bottomrule
\end{tabular}
}
\label{MASS}
\end{table}


\section{Baseline Methods}
We compare the proposed model with the following baselines:  

\begin{itemize}
\item  \textbf{Ref[2]}: Support vector machine is a typical machine learning method for classification tasks.
\item \textbf{Ref[3]}:  Random forests are an ensemble learning method for classification tasks.
\item 
\textbf{Ref[11]}: A mixed neural network, which combines multilayer perceptron (MLP) and LSTM. 
\item 
\textbf{Ref[12]}: A mixed model combines CNN and BiLSTM to capture time-invariant features and transition rules among sleep stages.
\item 
\textbf{Ref[7]}: A temporal sleep stage classification apply multivariate and multimodal time series.
\item 
\textbf{Ref[13]}: SeqSleepNet changes the single sleep stage classification problem into a sequence-to-sequence classification problem by using attention-based bidirectional RNN (ARNN) and RNN.

\item
\textbf{Ref[15]}: A spatial-temporal graph neural network with the adaptive sleep graph learning mechanism achieves the state-of-the-art performance. 
\end{itemize}

\section{Confusion matrices}

\begin{figure}[H]
    \centering
   \includegraphics[width=70mm]{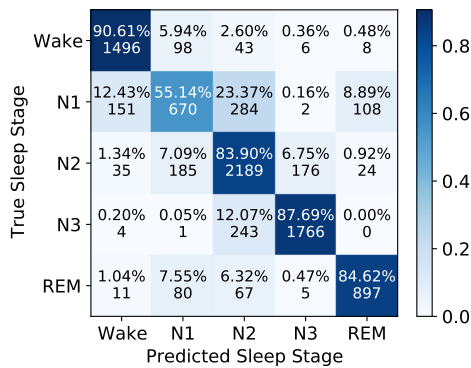}
   \caption{The confusion matrix of MSTGCN on ISRUC-S3 dataset.}
    \label{confusion matrix ISRUC}
\end{figure}

\begin{figure}[H]
    \centering
    \includegraphics[width=70mm]{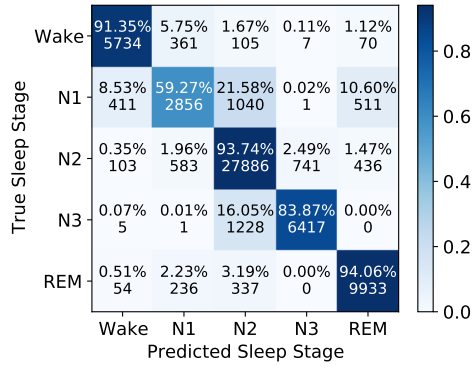}
    \caption{The confusion matrix of MSTGCN on MASS-SS3 dataset.}
    \label{confusion matrix MASS}
\end{figure}

Figure \ref{confusion matrix ISRUC} and Figure \ref{confusion matrix MASS} respectively present the confusion matrix of MSTGCN on different datasets. 

\section{Comparison of hypnograms scored by MSTGCN and sleep experts}

\begin{figure}[H]
\centering
\subfigure[Hypnogram manually scored by sleep experts.]{
\includegraphics[width=\textwidth]{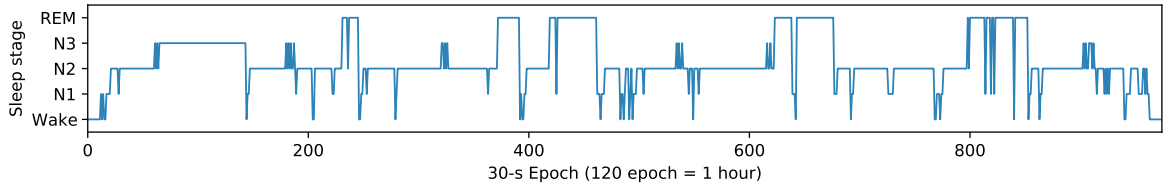}}
\subfigure[Hypnogram automatically scored by the proposed MSTGCN.]{
\includegraphics[width=\textwidth]{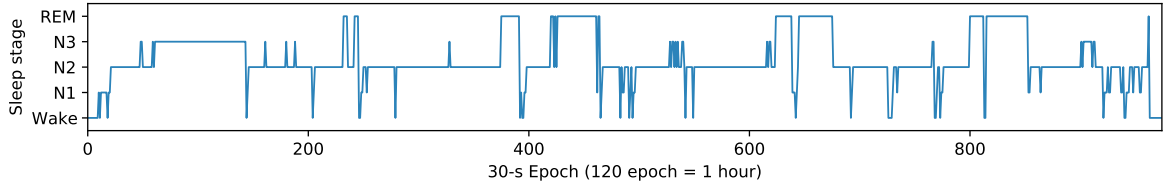}}
\caption{Output hypnogram produced by the proposed MSTGCN compared to the ground-truth. The hypnograms present that the proposed MSTGCN is able to classify most of the sleep stages correctly.} 
\label{Classification}
\end{figure}

\newpage

\section{Effect of Different Network Configurations}

\begin{figure}[htb]
    \centering
    \includegraphics[width=0.65\textwidth]{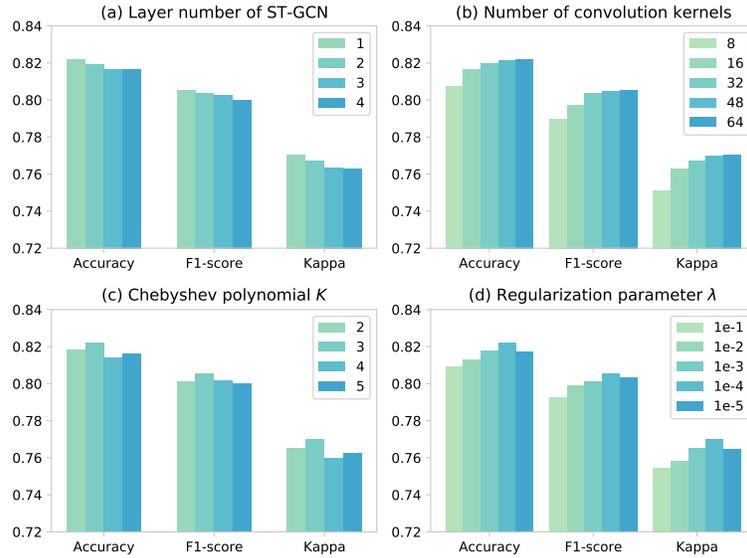}
    \caption{Network configuration analysis. The four figures respectively show the effect of layer number of ST-GCN, number of convolution kernels, Chebyshev polynomial $K$, and regularization parameter $\lambda$.}
    \label{Network-config}
\end{figure}

To further investigate how hyper-parameter settings affect the model performance, we conduct experiments with different network configurations. The same experiment hyperparameter setting is utilized in all experiments except the studied varying factor. As shown in Figure \ref{Network-config}, the proposed MSTGCN is not sensitive to the hyper-parameter settings. Fewer ST-GCN layers and more convolution kernels help to improve performance slightly. The appropriate Chebyshev polynomial $K$ and regularization parameter $\lambda$ is also conducive to the improvement of the model performance.

